\documentclass[twoside,slac_one]{revtex4}
\usepackage{graphicx}
\usepackage{fancyhdr}
\usepackage{amsmath} 
\usepackage{bm}
\usepackage{amsxtra}
\usepackage{amssymb}
\usepackage{amsthm}
\usepackage{latexsym}
\usepackage{lscape}

\begin{document}

\title{QCD: Experimental Review}

%

\author{J. Huston}
\affiliation{Department of Physics and Astronomy, Michigan State University, East Lansing, MI 48840}

\begin{abstract}
In this talk, I review progress in experimental QCD in the last year, concentrating on the results and phenomenology of the first year of running of the LHC. 

\end{abstract}

\maketitle

\thispagestyle{fancy}


\section{Introduction}

We are all looking for Beyond the Standard Model physics at the LHC. But before we publish any such discoveries, we need to make sure that we measure and understand Standard Model cross sections. This largely means understanding QCD, both perturbative and non-perturbative, at the LHC, in final states involving vector bosons, jets, photons and heavy quarks. So 2010 was largely the year of the Re-discovery of the Standard Model at the LHC~\footnote{My phrase, by the way.}, and in this talk, I will review some of the most important aspects of this re-discovery. 

\section{PDFs and LHC W and Z cross sections}
Crucial to the understanding of cross section predictions at the LHC is the knowledge of parton distribution functions (PDFs) and their uncertainties. The PDF4LHC group~\cite{PDF4LHC_website} was formed in order to explore the similarities and differences among the 6 groups that produce PDFs at next-to-leading (NLO) and next-to-next-to-leading order (NNLO). A benchmarking exercise was carried out to which all PDF groups were invited to participate~\cite{PDF4LHC_snapshot}. This exercise considered only the-then most up to date published versions/most commonly used of NLO PDFs from  6 groups: ABKM/ABM~\cite{Alekhin:2009ni, Alekhin:2010iu}, CTEQ/CT (CTEQ6.6~\cite{Nadolsky:2008zw}, CT10~\cite{Lai:2010vv}),GJR~\cite{Gluck:2007ck,Gluck:2008gs}, HERAPDF (HERAPDF1.0~\cite{herapdf10}), MSTW (MSTW08~\cite{Martin:2009iq}), NNPDF (NNPDF2.0~\cite{Ball:2010de}).
The benchmark cross sections were evaluated at NLO at both 7 and 14 TeV, but only the 7 TeV results have been published so far.

To perform a meaningful comparison, 
it is useful to first introduce the idea of  differential parton-parton luminosities. Such luminosities, when multiplied by the dimensionless 
cross section $\hat{s}\hat{\sigma}$ for a given process, provide a useful estimate  of 
the size of an event cross section at the LHC. 
The differential parton-parton luminosity
$dL_{ij}/d\hat{s}$ is defined as:

\begin{equation}
\frac{d L_{ij}}{d\hat{s}\,dy} = 
\frac{1}{s} \, \frac{1}{1+\delta_{ij}} \, 
[f_i(x_1,\mu) f_j(x_2,\mu) + (1\leftrightarrow 2)] \; .
\label{eq1}
\end{equation}
The prefactor with the Kronecker delta avoids double-counting in case the
partons are identical.  The generic parton-model formula 
\begin{equation}
\sigma = \sum_{i,j} \int_0^1 dx_1 \, dx_2 \, 
f_i(x_1,\mu) \, f_j(x_2,\mu) \, \hat{\sigma}_{ij}
\end{equation}
can then be written as 
\begin{equation}
\sigma = \sum_{i,j} \int \left(\frac{d\hat{s}}{\hat{s}} \right) 
\, \left(\frac{d L_{ij}}{d\hat{s}}\right) \, 
\left(\hat{s} \,\hat{\sigma}_{ij} \right) \; .
\label{eq:xseclum}
\end{equation}

\begin{figure}
\begin{center}
\includegraphics[width=0.48\textwidth]{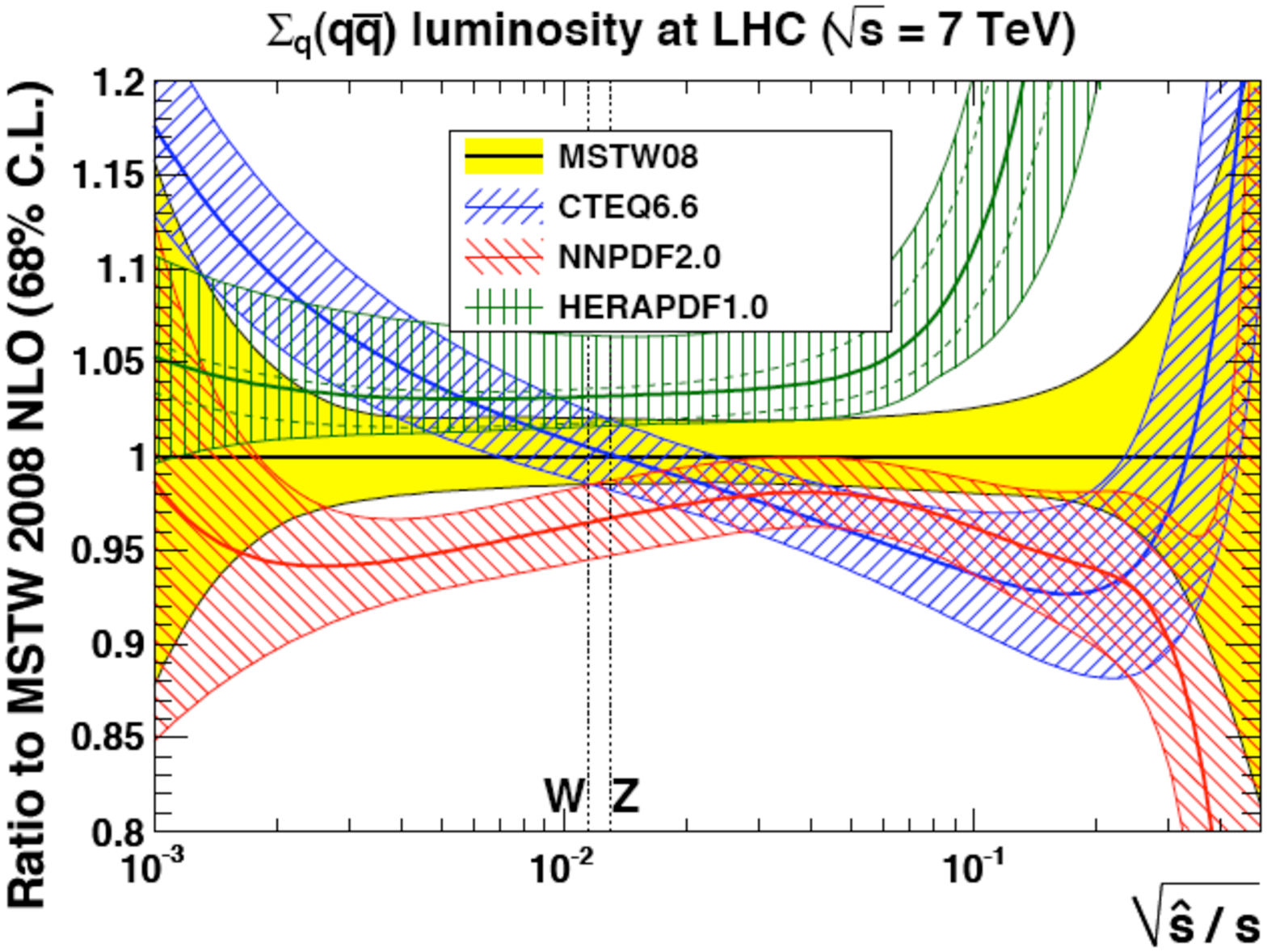}
\includegraphics[width=0.48\textwidth]{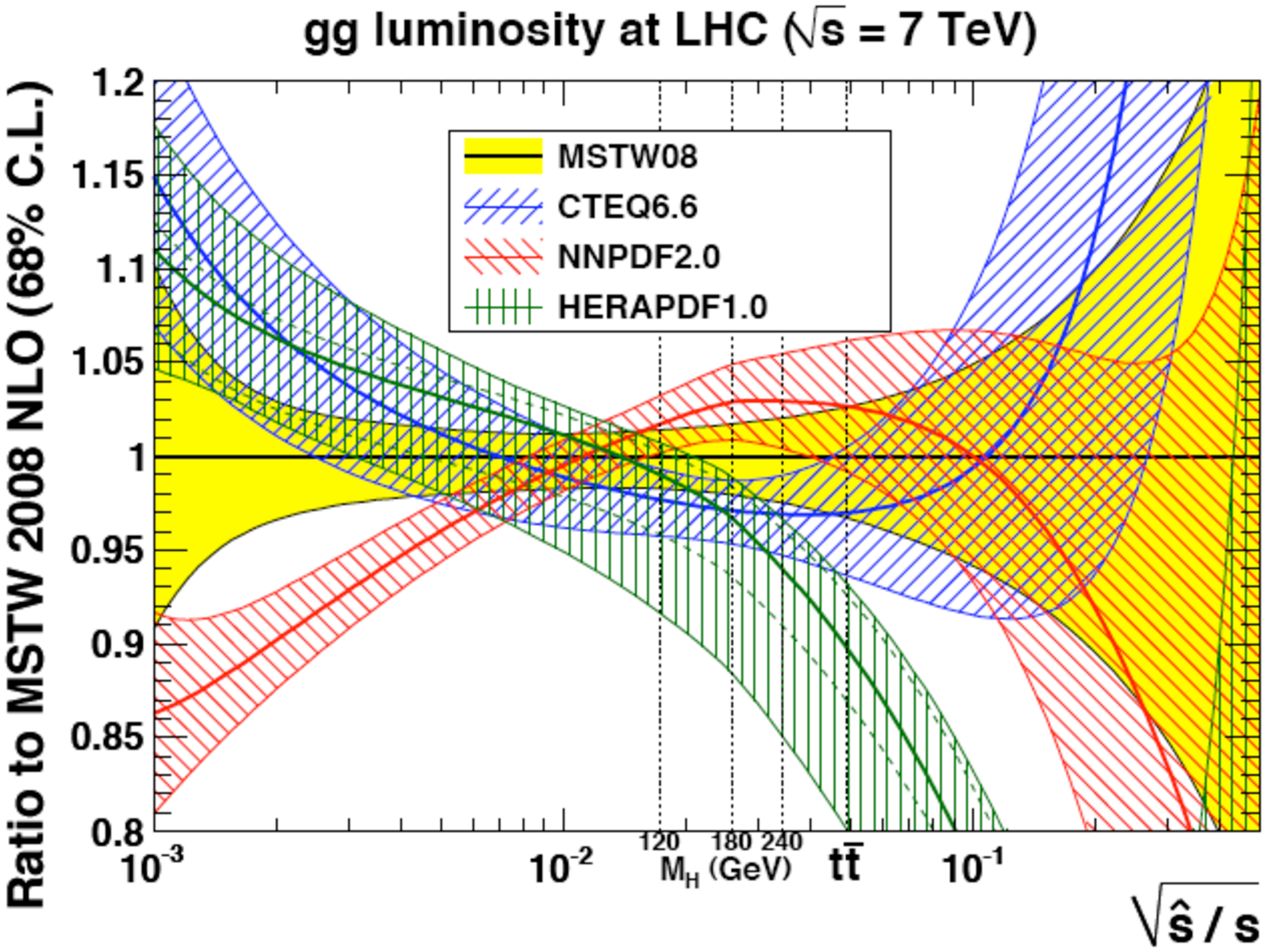}
\caption{\label{fig:qqlum} The $q\bar{q}$ (left) and $gg$ (right) luminosity functions and
  their uncertainties at 7 TeV, normalized to the MSTW08 result. 
Plot by G. Watt \cite{Watt}.}
\end{center}
\end{figure}

Relative quark-antiquark and gluon-gluon PDF
luminosities are shown in Figure~\ref{fig:qqlum}.
CTEQ6.6,
NNPDF2.0, HERAPDF1.0 and MSTW08 PDF luminosities are shown (the ABKM09 and GJR08 comparisons are available in the PDF4LHC writeup), all
normalized to the MSTW08 central value,
along with their 68\%c.l.
error bands. The inner uncertainty bands (dashed lines) for HERAPDF1.0 
correspond to the (asymmetric) experimental errors, while the
outer uncertainty bands (shaded regions) also include the model and
parameterisation errors. It is
interesting to note that the error bands for each of the PDF
luminosities are of  similar size, even though different criteria are often used to determine the tolerance in the global fits. The
predictions of W/Z, $t\bar{t}$ and Higgs cross sections  are in
reasonable agreement for CTEQ, MSTW and NNPDF, while the agreement
with ABKM, HERAPDF and GJR is somewhat worse.
(Note however that these plots do not illustrate the effect that 
the different $\alpha_s(m_Z)$ values used by different groups will have 
on (mainly) $t\bar{t}$ and Higgs cross sections.)
It is also notable that the PDF
luminosities tend to differ at low $x$ and high $x$, for both
$q\bar{q}$ and $gg$ luminosities. The CTEQ6.6 distributions, for
example, may be larger at low $x$ than MSTW2008, due to the
positive-definite parameterization of the gluon distribution; the MSTW
gluon starts off negative at low $x$ and $Q^2$ and this results in an
impact for both the gluon and sea quark distributions at larger $Q^2$
values. The NNPDF2.0 $q\bar{q}$ luminosity tends to be somewhat lower,
in the $W,Z$ region for example.  Part of this effect might come from
the use of a ZM heavy quark scheme, although other differences might 
be relevant (subsequent  versions of NNPDF use a GM heavy quark scheme, similar to the other PDF groups). 

\begin{figure}
\begin{center}
\includegraphics[width=0.48\textwidth]{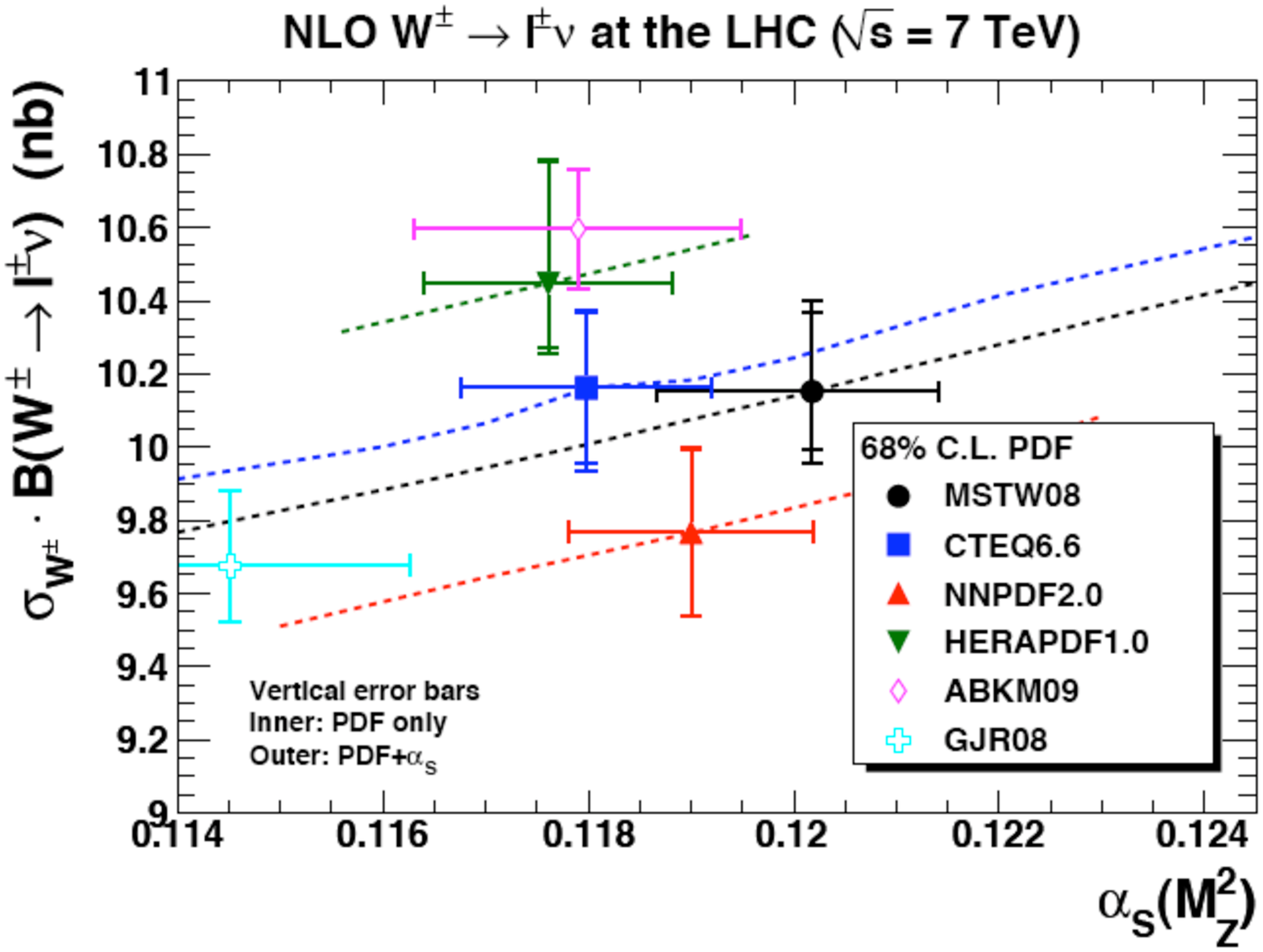}
\includegraphics[width=0.48\textwidth]{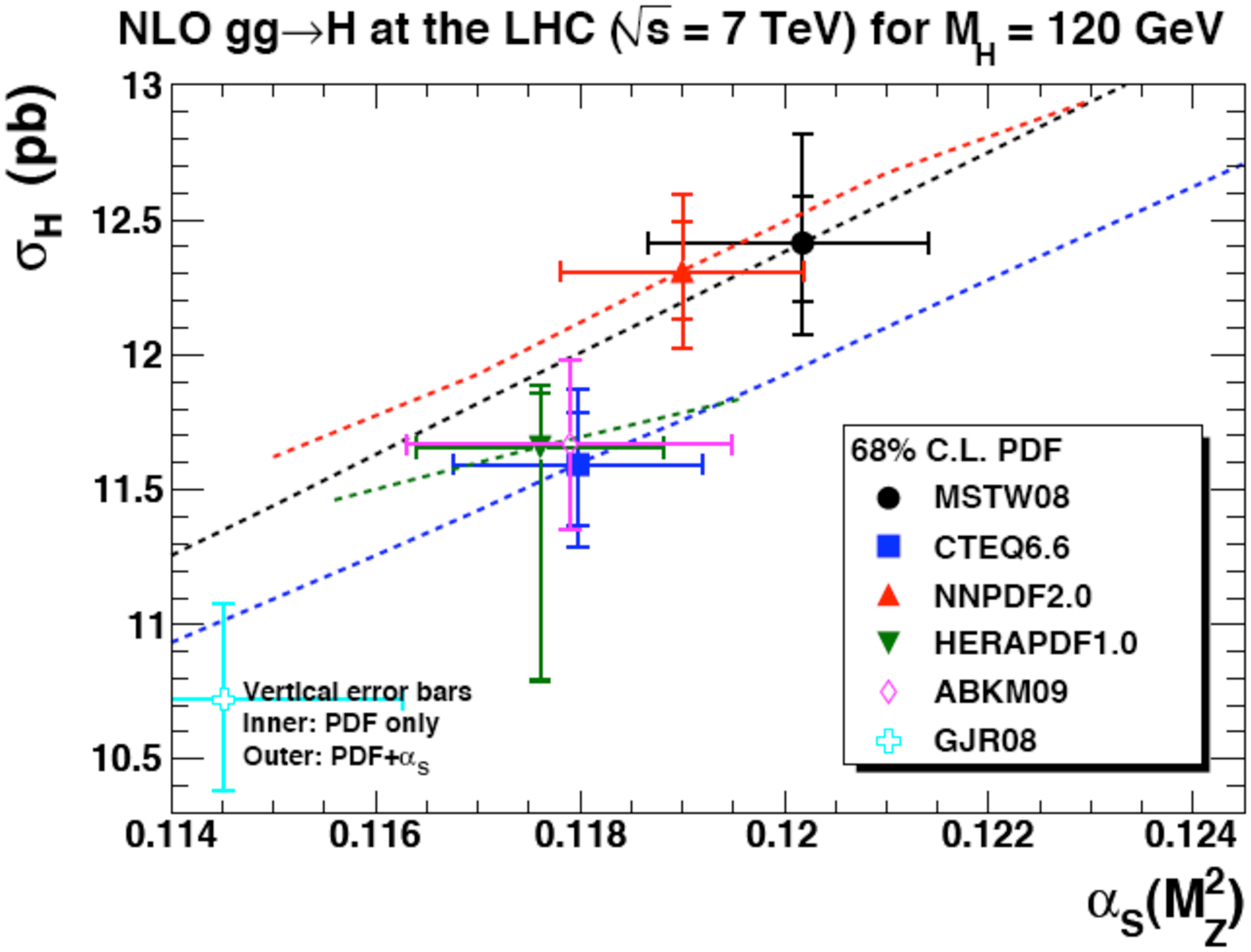}
\caption{\label{fig:WZ1} Cross section predictions at 7 TeV for $W^\pm$ and $Z$ production (left) and Higgs production ($m_H$=120 GeV) through $gg$ fusion (right). All $Z$ cross sections plotted here
use a value of $\sin^2\theta_W=0.23149$. Plot by G. Watt \cite{Watt}.}
\end{center}
\end{figure}

After having performed the comparison between PDF luminosities, it is now useful to compare predictions for LHC observables. Perhaps the best manner
to perform this comparison is to show the cross--sections as a
function of $\alpha_s$, with an interpolating curve connecting
different values of $\alpha_s$ for the same group, when
available~\cite{Watt} (see
Fig.~\ref{fig:WZ1}). Following the interpolation
curve, it is possible to compare
cross sections at the same value of $\alpha_s$.
The predictions for the CTEQ, MSTW and NNPDF $W$ and $Z$ cross
sections at 7 TeV (Fig.~\ref{fig:WZ1})
agree well, with  the NNPDF predictions somewhat lower, consistent
with the behaviour of the PDF luminosities observed in Fig.~\ref{fig:qqlum}.
The impact from the variation of the value of $\alpha_s$ is
relatively small. Basically, all of the PDFs
predict similar values for the $W/Z$ cross section ratio; much of the remaining uncertainty in this ratio is related to uncertainties in the strange quark distribution. This ratio will also
serve as a useful benchmark at the LHC.

The predictions for Higgs production  from $gg$ fusion
(Figs.~\ref{fig:WZ1}) depend strongly on the value
of $\alpha_s$: the anticorrelation between the gluon distribution and
the value of $\alpha_s$ is not sufficient to offset the growth of the
cross section (which starts at $O(\alpha_s^2)$ and undergoes a large
$O(\alpha_s^3)$  correction).
The CTEQ, MSTW and NNPDF predictions are in moderate agreement
but CTEQ lies somewhat lower, to some extent due to
the lower choice of $\alpha_s(M_Z^2)$. Compared at the common value of 
$\alpha_s(M_Z^2)=0.119$, the CTEQ prediction and that of either MSTW or
NNPDF, have one-sigma PDF uncertainties which just about overlap for each
value of $m_H$. If the comparison is made at the respective reference 
values of $\alpha_s$, but without accounting for the $\alpha_s$ uncertainty,
the discrepancies are rather worse, and indeed, even allowing for $\alpha_s$
uncertainty, the bands do not overlap. Hence, both the difference in PDFs and
in the dependence of the cross section on the value of $\alpha_s$ are
responsible for the differences observed. 

\subsection{PDF4LHC interim prescription}

One of the charges of the PDF4LHC group is to provide a protocol for both experimentalists and theorists to use PDF sets to  calculate central cross sections, and uncertainties, at the LHC. There are two separate recommendations, one for NLO cross sections and one for NNLO cross sections~\cite{PDF4LHC_recommend}. As observed previously, there is reasonable agreement among the predictions from CTEQ, MSTW and NNPDF, with somewhat larger deviations from the other PDF groups. However, the error estimate from any single PDF group does not cover the full range of predictions from all three PDFs with their uncertainties. Thus the NLO prescription is to use the envelope provided by the central values and PDF$+\alpha_s$ errors from the MSTW08, CTEQ6.6 and NNPDF2.0 PDFs, using each group's prescriptions for combining the two types of errors. The prescription at NNLO is to base the predictions on MSTW08 at NNLO, since that was the only NNLO PDF based on a fully global PDF fit at the time that the prescription was codified (as mentioned previously now all 6 PDF groups have PDFS available at NNLO). But, since there seems to be no reason to believe that the spread in predictions of the global fits will diminish at NNLO compared to NLO, the NNLO uncertainty from MSTW08 should be expanded by the ratio of the NLO envelope to the uncertainty from MSTW08 NLO alone. This basically results in a doubling of the MSTW08 NNLO uncertainty. 

Note that this prescription is most useful when calculating cross sections not yet measured (as for example Higgs production), or for a more conservative estimate of the acceptance uncertainties for a particular cross section. For comparison to existing experimental cross sections, it is better to provide direct comparisons with individual PDF predictions; such cross section measurements have the potential to feed back into the global fits and to modify the resulting new PDFs. 

The benchmarking exercise (and accompanying recommendations) were published at the beginning of 2011. There are new data from HERA, from the Tevatron and from the LHC, as well as new PDFs from the fitting groups, including for the first time NNLO PDFs from all 6 groups. The exercise is being updated to take these developments into account. 

\section{W/Z production}

\begin{figure}
\begin{center}
\includegraphics[width=0.40\textwidth]{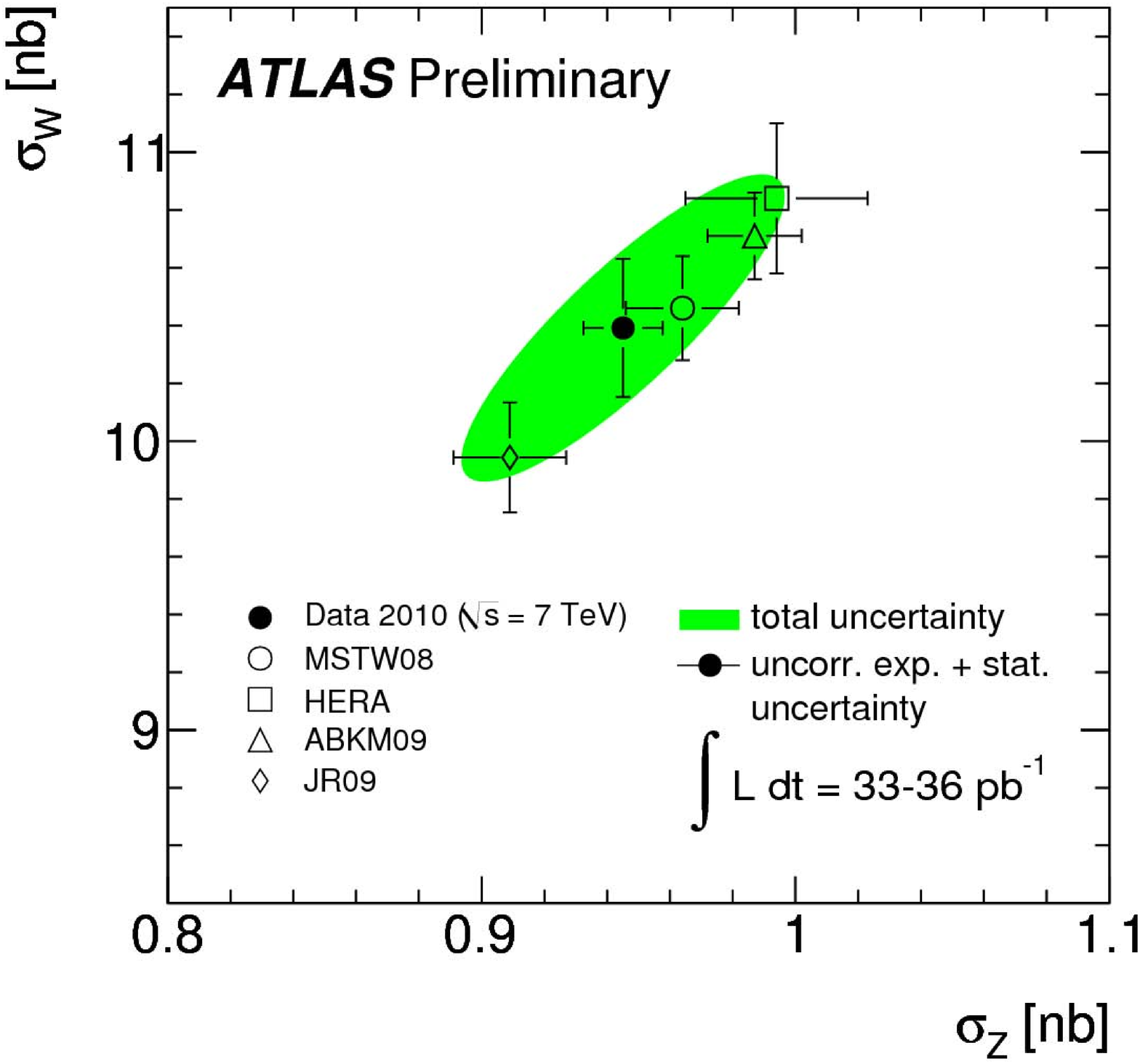}
\includegraphics[width=0.56\textwidth]{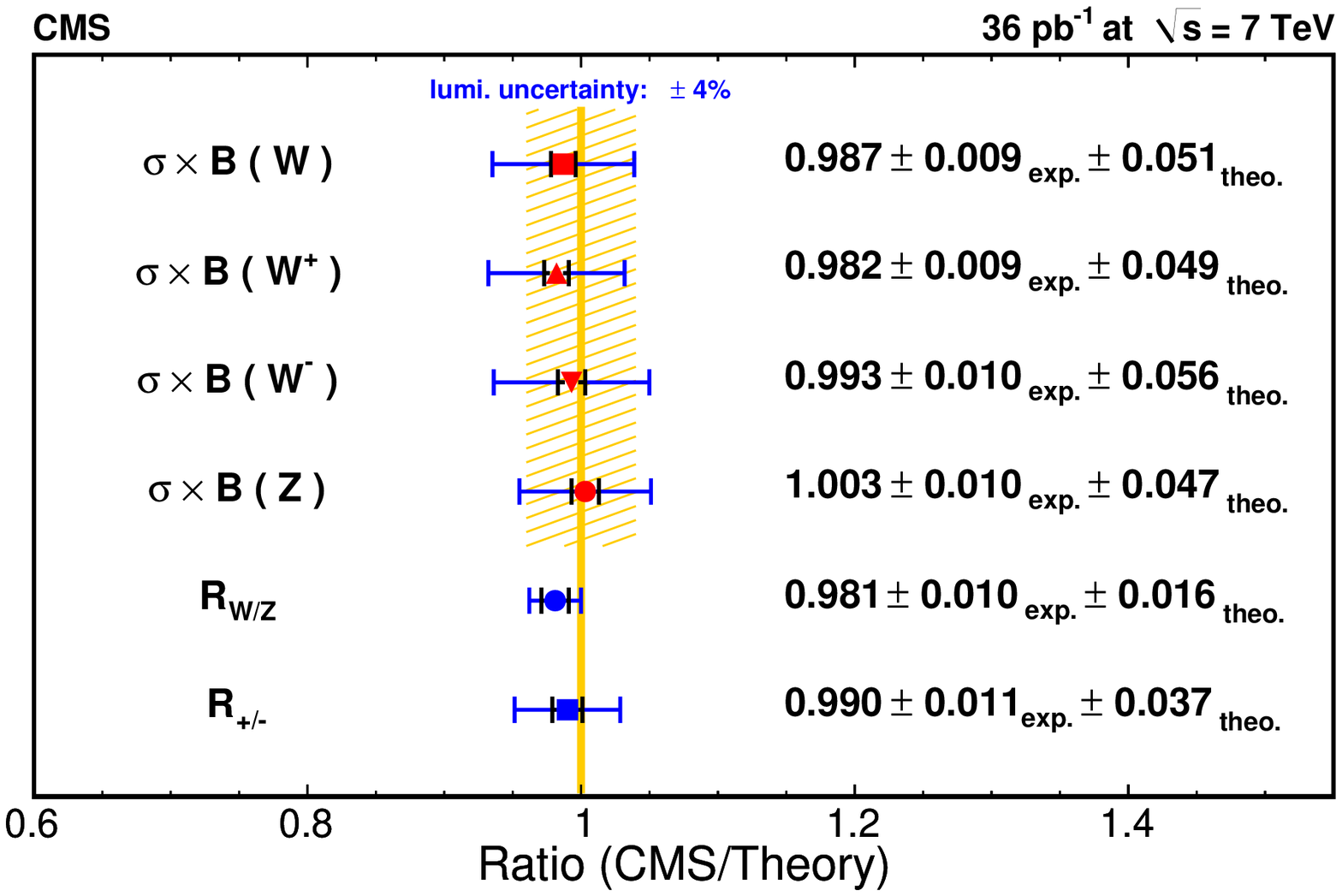}
\caption{(left) The measured and predicted (at NNLO) cross sections times leptonic branching ratios for ATLAS. The projections of the ellipse to the axes correspond to a 1-sigma uncertainty of the cross sections. (right) The ratios of the measured cross sections to the theoretical predictions (at NNLO) for CMS. The predicted cross sections and uncertainty ranges have been derived using the PDF4LHC interim prescription.  \label{fig:WZ_ATLAS_CMS} }
\end{center}
\end{figure}

A comparison of the measured $W$ and $Z$ cross sections to NNLO predictions for ATLAS~\cite{ATLASW} (left) and CMS~\cite{CMSW} (right) are shown in Fig.~\ref{fig:WZ_ATLAS_CMS}. In general, the predictions are in good agreement with the data from both experiments. The ATLAS measurements are compared to NNLO predictions using the specific PDFs shown in the figure. The CMS data are compared to a NNLO prediction and uncertainty determined using the PDF4LHC prescription. 
Boson production will continue to serve as the primary benchmark for Standard Model cross sections at the LHC.

\section{The Underlying Event at 900 GeV and 7 TeV}

\begin{figure}
\begin{center}
\includegraphics[width=0.40\textwidth,angle=90]{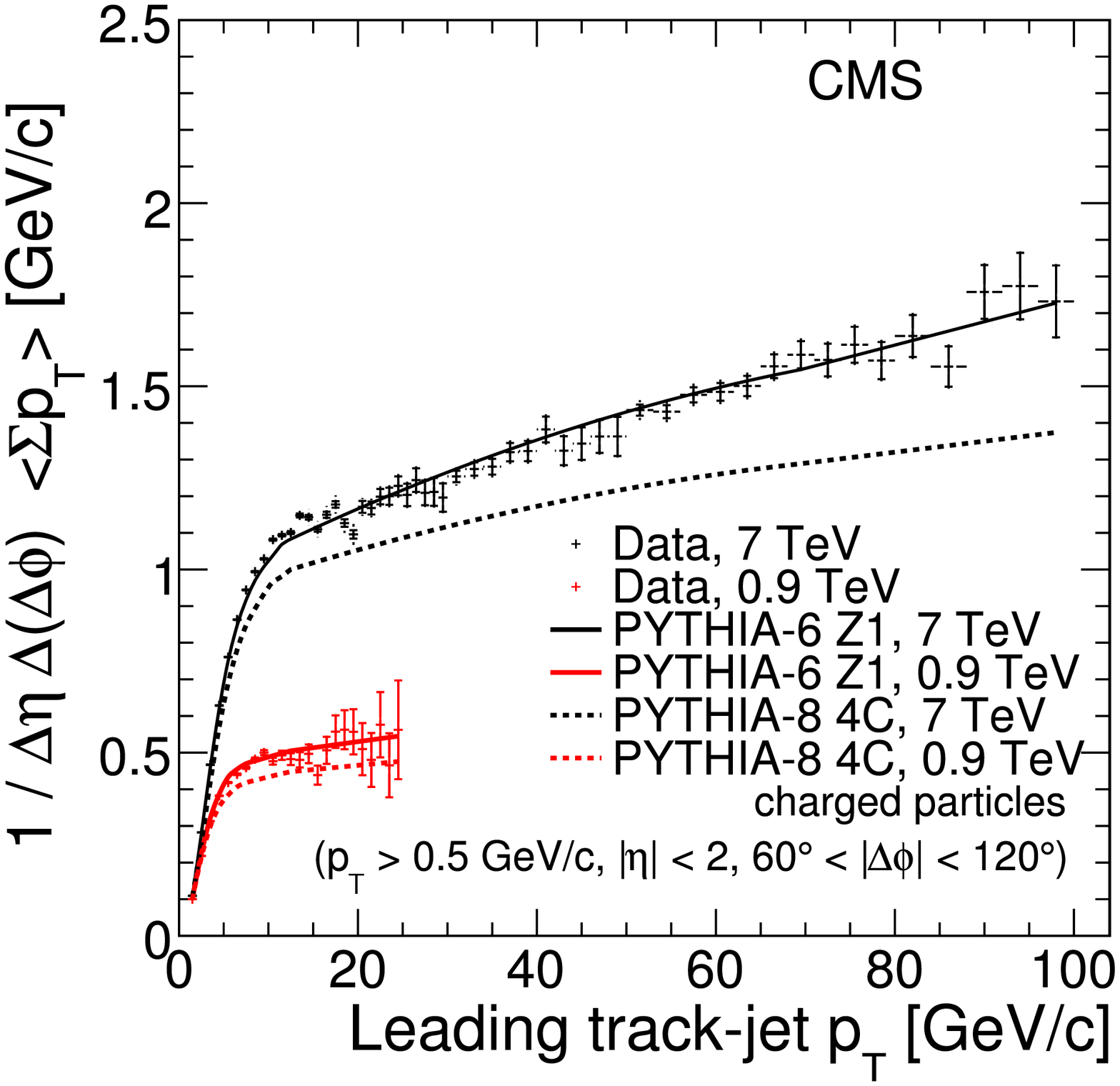}
\includegraphics[width=0.40\textwidth,angle=90]{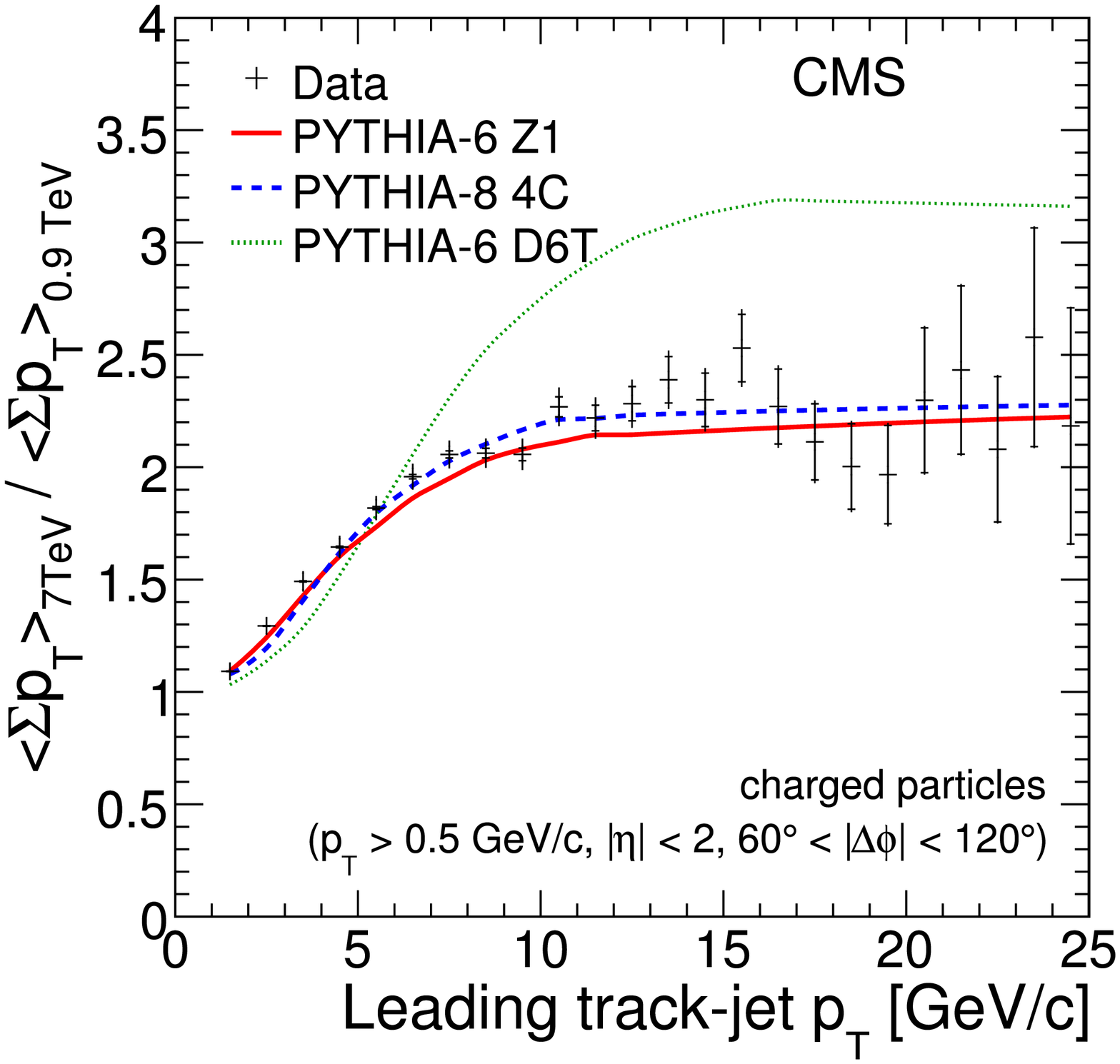}
\caption{(left) The charged particle $\Sigma p_T$ in the transverse region in CMS at 900 GeV and 7 TeV. (right) The ratio of the   charged particle $\Sigma p_T$ distributions at the two energies. \label{fig:CMS_UE} }
\end{center}
\end{figure}

Critical to the achievement of precision physics at the LHC is the determination of the $\it Underlying$ $\it Event$ (UE). For inclusive jet physics, for example, the effects of the UE have to be accounted for in the comparison of the measured jet cross section to partonic level predictions; this can be accomplished either by subtracting the UE from the data, or adding the effects of the UE to the partonic theory~\cite{CHS}. The UE is created primarily by the semi-hard scatters (of the order of a few GeV) of low $x$ gluons. As the gluon distribution rapidly increases as $x$ goes to zero, the UE was expected to increase at the LHC, compared to the level observed at the Tevatron. Both ATLAS~\cite{ATLAS_UE} and CMS~\cite{CMS_UE} have carried out extensive studies of the UE. For example, the sum of the charged particle transverse momenta (for $p_T > 500 MeV, |\eta| < 2, 60^o < |\Delta \Phi| < 120^o$ with respect to the leading jet) is shown for both 900 GeV and 7 TeV, for CMS, in Fig.~\ref{fig:CMS_UE}. A strong growth is observed from 900 GeV to 7 TeV that can be well-described by  parton shower Monte Carlo predictions with the appropriate tunes. There is still a difficulty, however, in simultaneously describing the Tevatron data as well. 

\section{Jet production}

Both ATLAS~\cite{ATLASjet} and CMS~\cite{CMSjet} have carried out extensive investigations of inclusive jet and dijet production over a very wide kinematic range. Two pieces of good news are that (1) both experiments are using the antikT jet algorithm~\cite{antikT}, an IR-safe jet algorithm and (2) that both experiments are carrying out these measurements with two separate jet sizes. The bad news is that ATLAS is using jet sizes of 0.4 and 0.6, while CMS is using jet sizes of 0.5 and 0.7, i.e. there are no common sizes. Hopefully, this lack of commonality will be rectified in the future. Both experiments have the capability of using multiple jet algorithms/sizes in their analyses without the extensive re-calibration necessary at the Tevatron, and the utilization of such flexibility is essential if the perturbative/non-perturbative environment at the LHC is to be truly explored~\cite{CHS}. For example, the influence of underlying event/pileup decreases for smaller jet sizes, but the non-perturbative fragmentation effects become larger~\cite{gavin}. In addition, as $R_{jet}$ decreases, $lnR$ terms in the perturbative expansion may become important. But as $R_{jet}$ increases, there can also be an increase in the scale uncertainty for the resultant cross section due to the inclusion of more tree-level gluon radiation. 

\begin{figure}
\begin{center}
\includegraphics[width=0.48\textwidth]{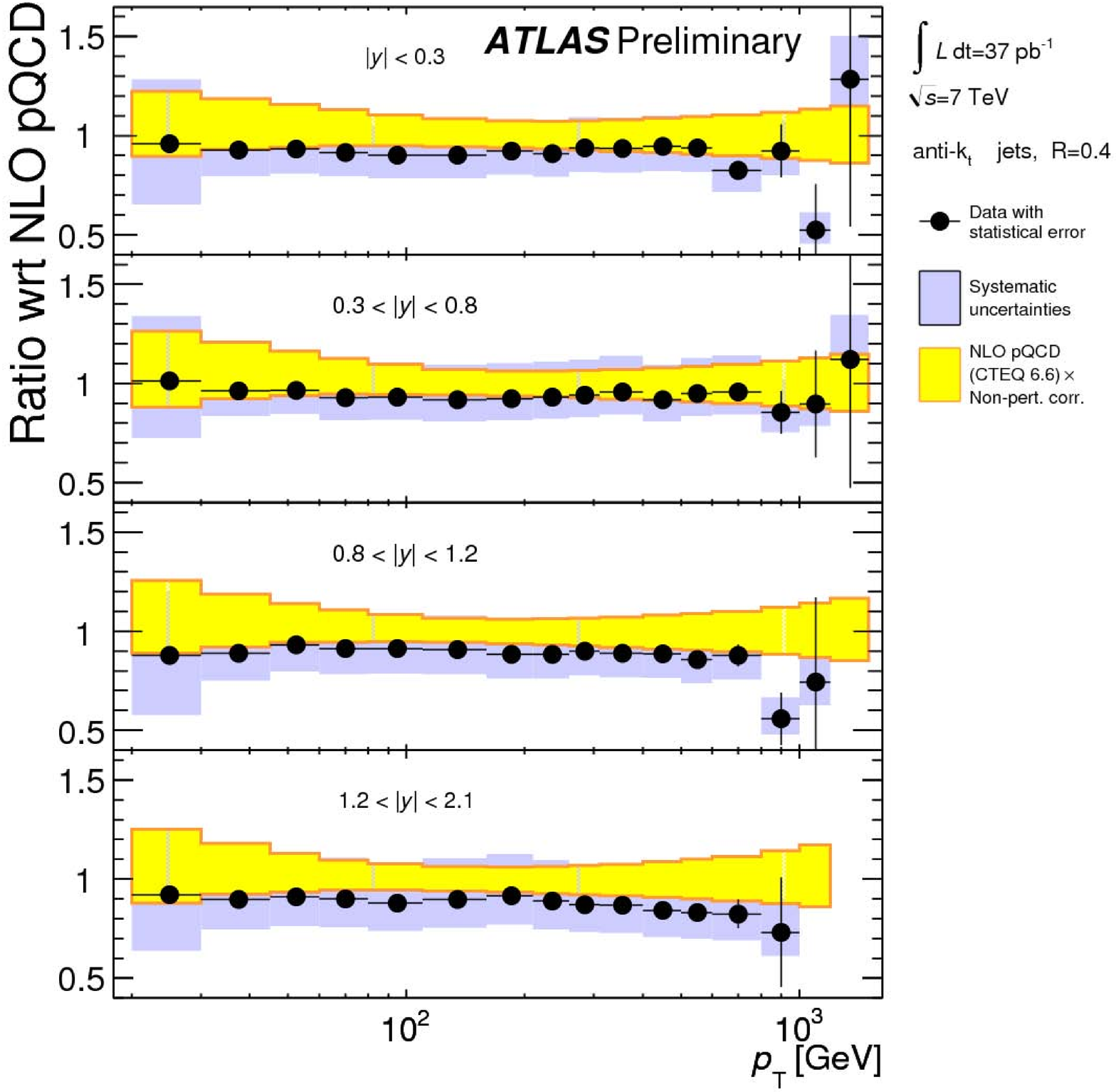}
\includegraphics[width=0.48\textwidth]{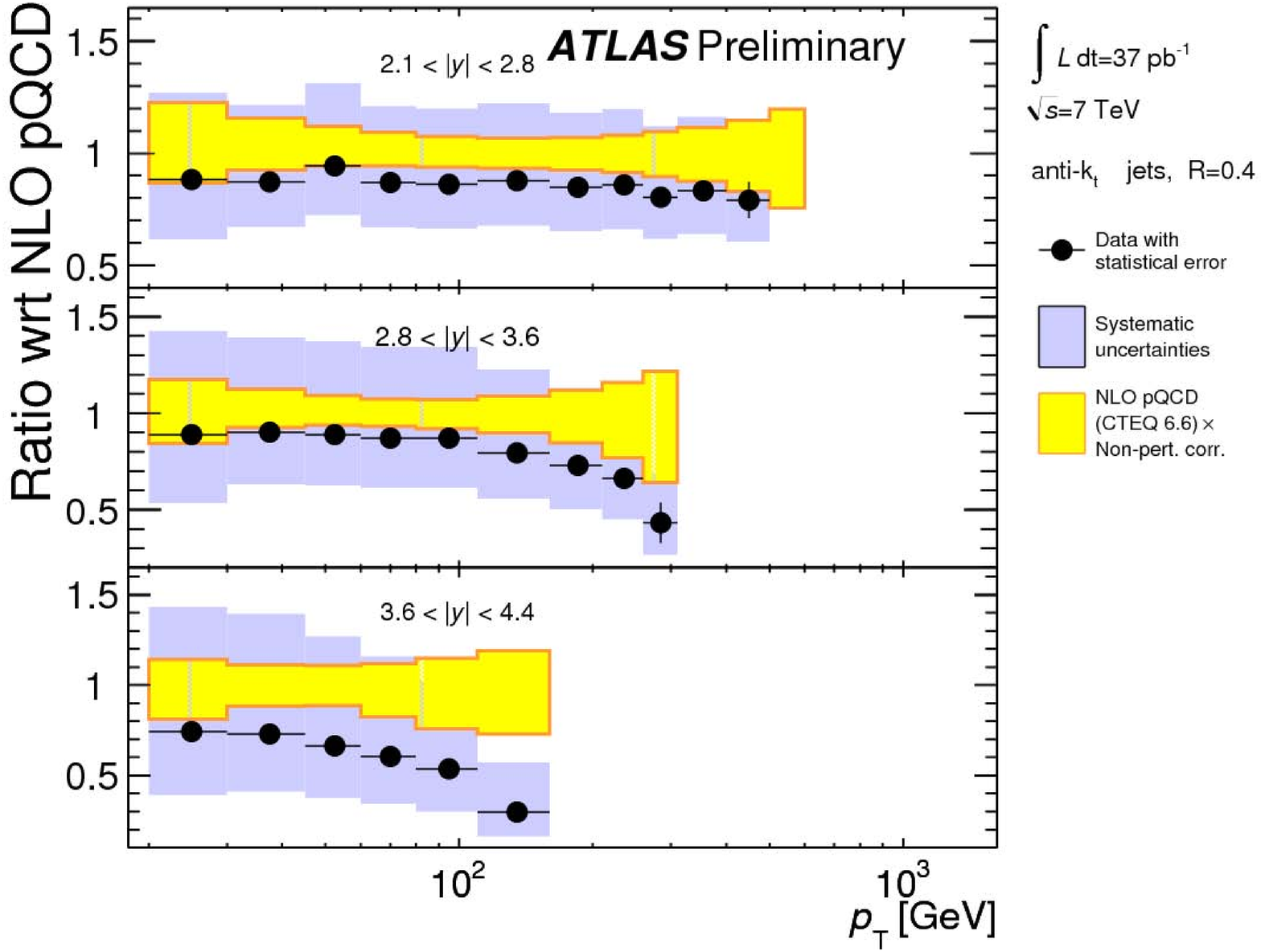}
\caption{The inclusive differential jet cross section is plotted as a function of jet $p_T$ in different regions of 
$|y|$ for jets reconstructed with the antikT algorithm with radius R=0.4 (similar plots exist for R=0.6). The ratio of the data to the NLO theoretical predictions (corrected for underlying event and non-perturbative fragmentation) is shown, along with the total systematic uncertainties for data and theory. \label{fig:ATLAS_jets} }
\end{center}
\end{figure}

A comparison of the ATLAS inclusive jet cross section to NLO predictions~\cite{NLOJET++} using several PDFs is shown in Fig.~\ref{fig:ATLAS_jets}. A renormalization and factorization scale of $p_T^{max}$, the transverse momentum of the leading jet in the event, has been used; variations of a factor of 2 higher and lower than that value are used to estimate the theoretical uncertainties from uncalculated higher order corrections. Note that the measurement is made over a wide rapidity range; this is crucial if possible new physics is to be separated from more mundane old physics, such as lack of knowledge of PDFs at low/high x~\cite{CTEQ6.1}. There is reasonable agreement with the PDFs with which the comparisons are made. Note that for a detailed check of the level of agreement, the correlated systematic errors are necessary. These correlated errors are also necessary for the inclusion of this data into global PDF fits. There is the tendency for the data to be lower than the predictions for high $p_T$ and $y$, a tendency also observed at the Tevatron in Run 2 (but not in Run 1). More detailed investigations possible with the 2011 data will be critical in furthering our understanding. 

\begin{figure}
\begin{center}
\includegraphics[width=100mm]{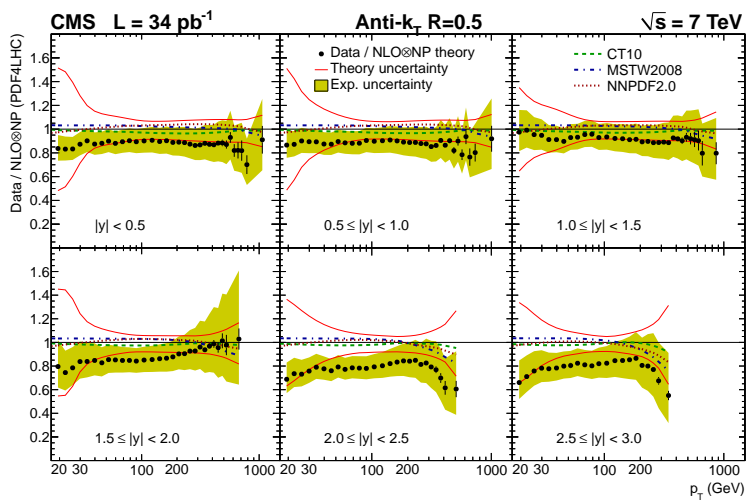}
\caption{The inclusive differential jet cross section is plotted as a function of jet $p_T$ in different regions of 
$|y|$ for jets reconstructed with the antikT algorithm with radius R=0.5 (similar plots exist for R=0.7). The ratio of the data to the NLO theoretical predictions (corrected for underlying event and non-perturbative fragmentation) is shown, along with the total systematic uncertainties for data and theory. The predictions have been carried out using the PDF4LHC interim recommendation and a factorization/renormalization scale equal to the transverse momentum of the highest jet in the event has been used. Central predictions for the CT10 (dashed line), MSTW2008 (dash-dotted line) and NNPDF (dotted line) PDF sets are also shown.  \label{fig:CMS_jets} }
\end{center}
\end{figure}

\begin{figure}
\begin{center}
\includegraphics[width=0.35\textwidth]{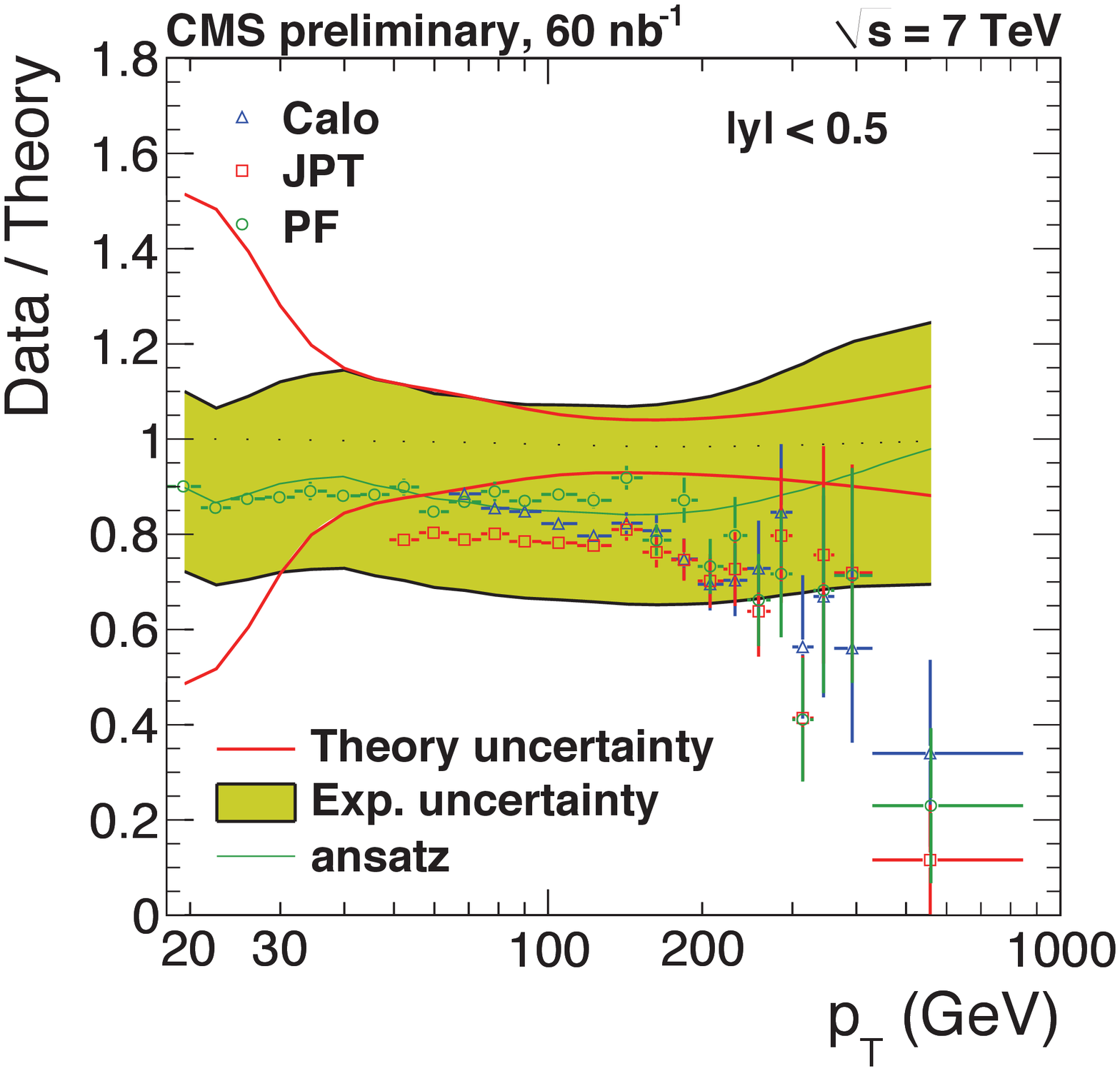}
\includegraphics[width=0.35\textwidth]{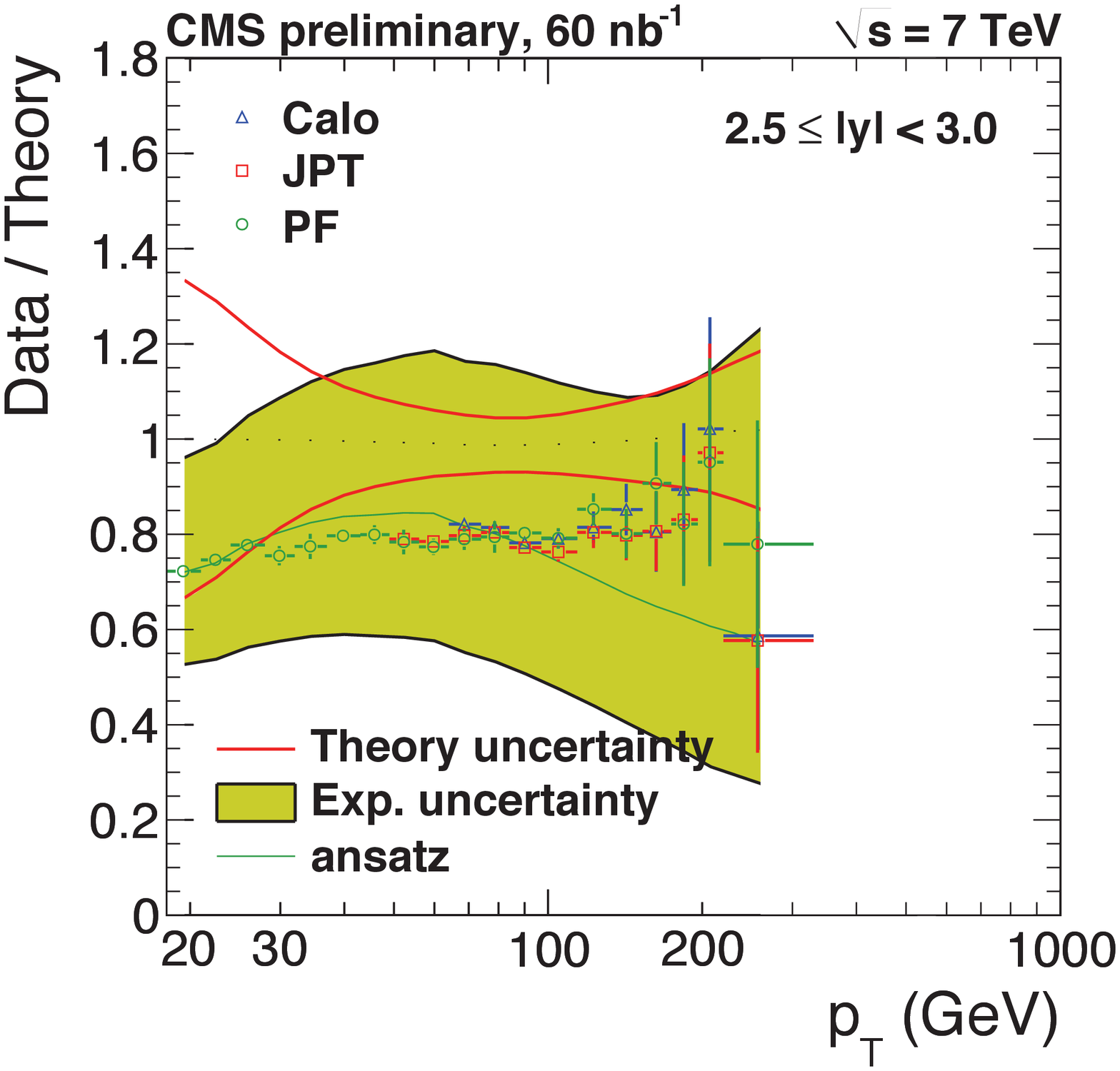}
\caption{The inclusive differential jet cross section is plotted as a function of the jet $p_T$ in two different rapidity regions, for jets reconstructed with the antikT algorithm with R=0.5. Three different methods of jet reconstruction are compared: calorimeter jets (triangles), JPT jets (squares) and particle-flow jets (circles). \label{fig:CMS_jet2} }
\end{center}
\end{figure}

A comparison of the CMS data with the antikT algorithm with R=0.5, over the full rapidity region is shown in Fig.~\ref{fig:CMS_jets}. There is a tendency for the data to be globally somewhat less than the NLO predictions, with some additional dropoff observed at high $p_T/y$. In Fig.~\ref{fig:CMS_jet2}, a comparison is made of the inclusive jet cross section observed in one of the central and one of the forward rapidity regions using three different techniques of measurement (calorimetric/jet plus tracks/particle flow). Such multiple techniques also serve to make the measurement more experimentally robust. 

\begin{figure}
\begin{center}
\includegraphics[width=0.48\textwidth]{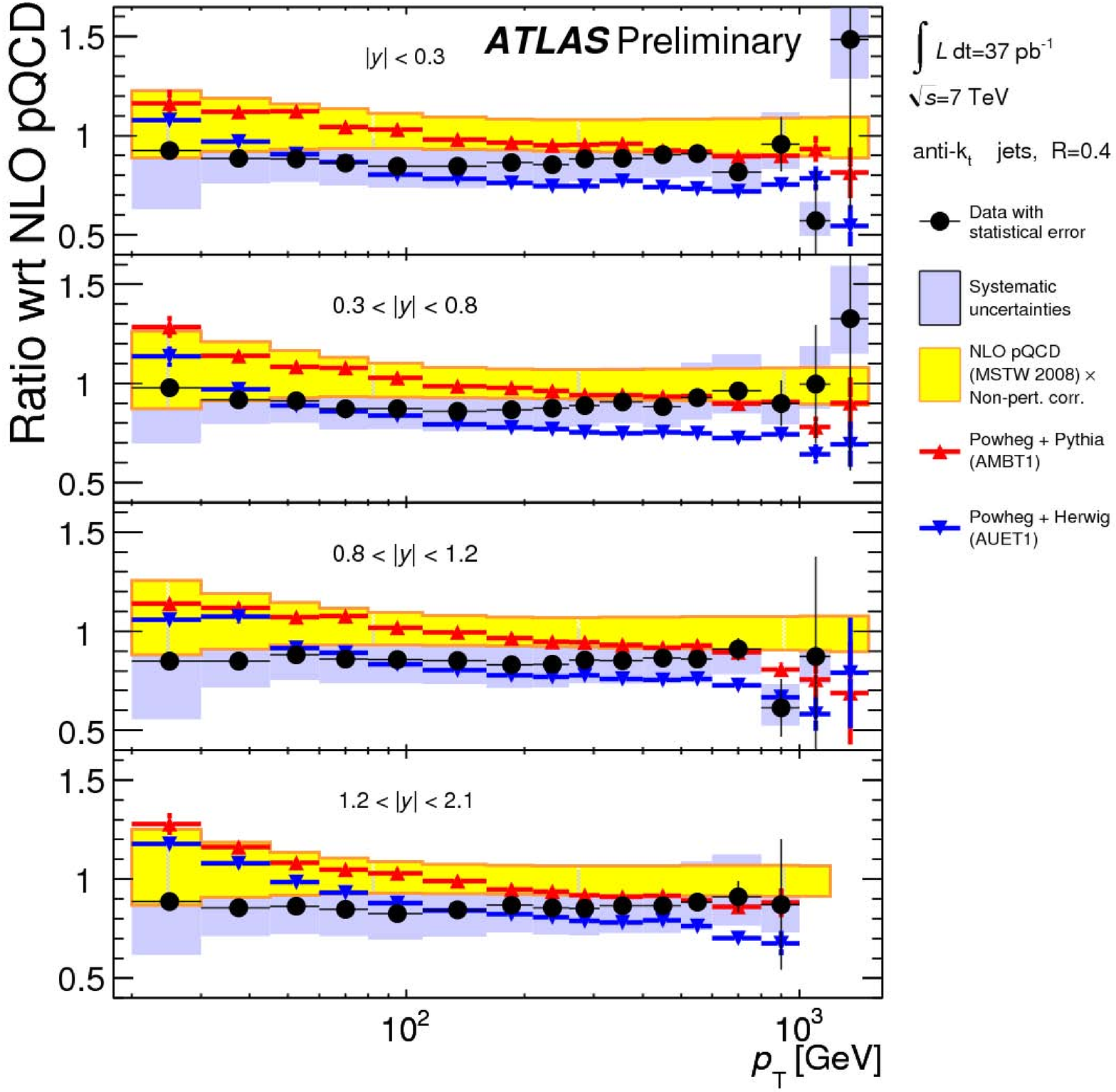}
\includegraphics[width=0.48\textwidth]{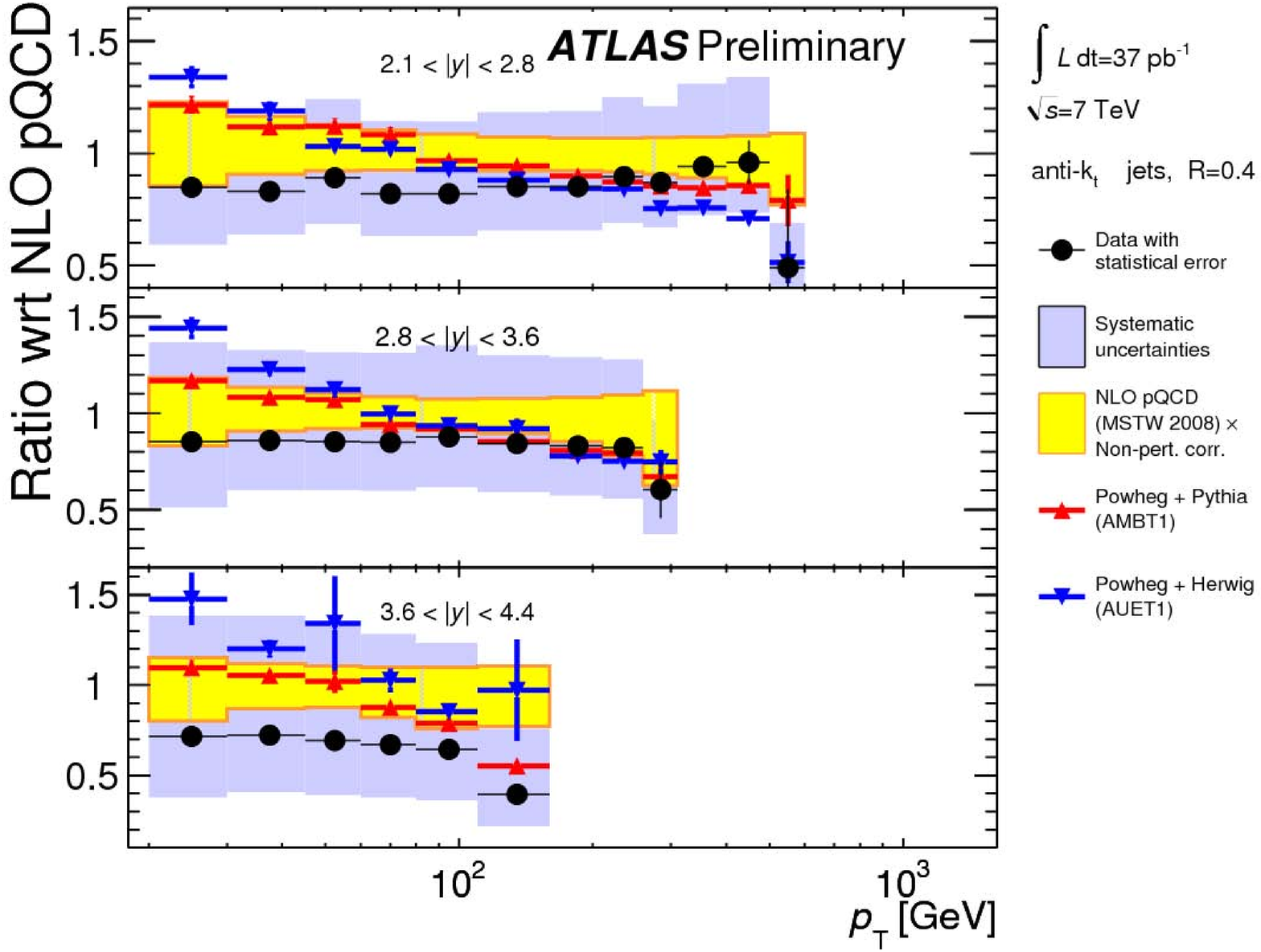}
\caption{The inclusive differential jet cross section is plotted as a function of jet $p_T$ in  different regions of 
$|y|$ for jets reconstructed with the antikT algorithm with radius R=0.4 (similar plots exist for R=0.6). The ratio of the data to the NLO theoretical predictions obtained with Powheg is shown, along with the total systematic uncertainties for data and theory. Results are shown for Powheg showered both with Pythia and with Herwig. The MSTW2008 PDFs have been used. \label{fig:ATLAS_Powheg} }
\end{center}
\end{figure}

In addition, ATLAS has compared its measurement of inclusive jet production to NLO+parton shower predictions from Powheg~\cite{Powheg}. See Fig.~\ref{fig:ATLAS_Powheg}. The predictions show some differences in detail from those from fixed order calculations (and indeed differences depending on whether Pythia~\cite{Pythia} or Herwig~\cite{Herwig} is used for the parton showering). Given the importance of the inclusive jet data to precision physics at the LHC, and specifically to PDF fits,  a better understanding of these differences is needed. Such studies are underway. 

ATLAS~\cite{ATLASjet} and CMS~\cite{cms_dijet} have also carried out studies of dijet production over a wide kinematic region, comparing the data to predictions from NLOJET++ (ATLAS and CMS) and from Powheg (ATLAS). Relatively good agreement with NLOJET++ is observed for the most part, with somewhat less of a tendency for the data to be smaller than the theoretical predictions near the kinematic edges, as compared to the comparisons for the inclusive jet cross section. For the ATLAS dijet comparisons to Powheg, there are differences in shape and normalization for the Powheg predictions compared to the fixed order predictions, similar to those observed for the inclusive jet case. 

The dijet cross section measurements at the LHC cover a very wide kinematic range, sampling both high and low $x$ parton distributions. There is an indication that larger values of the renormalization and factorization scales, than those that  work well in the central rapidity regions,  may be required to provide stable cross section predictions at large dijet mass and high $y_{max}$. Investigations are continuing. 


\section{W/Z + jets}

Final states involving vector bosons plus jets serve as a signal channel/background for both Standard Model (such as $t\bar{t}$ production) and non-Standard Model (such as supersymmetry) physics. Already, the data taken in 2010 allows for a new kinematic testing ground for the new theoretical predictions that have become available in the last several years~\cite{blackhat}. 

\begin{figure}
\begin{center}
\includegraphics[width=0.40\textwidth]{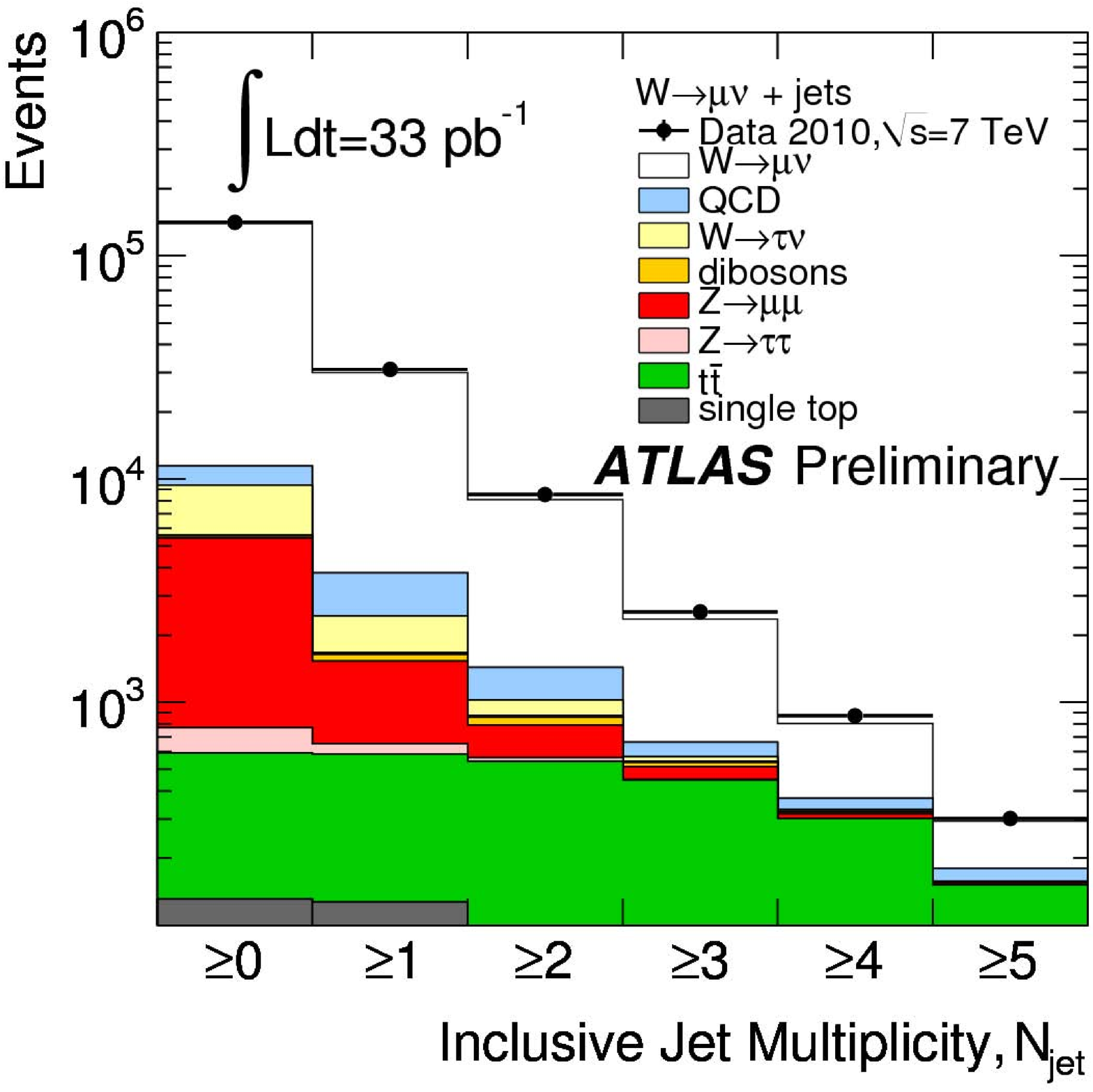}
\includegraphics[width=0.40\textwidth]{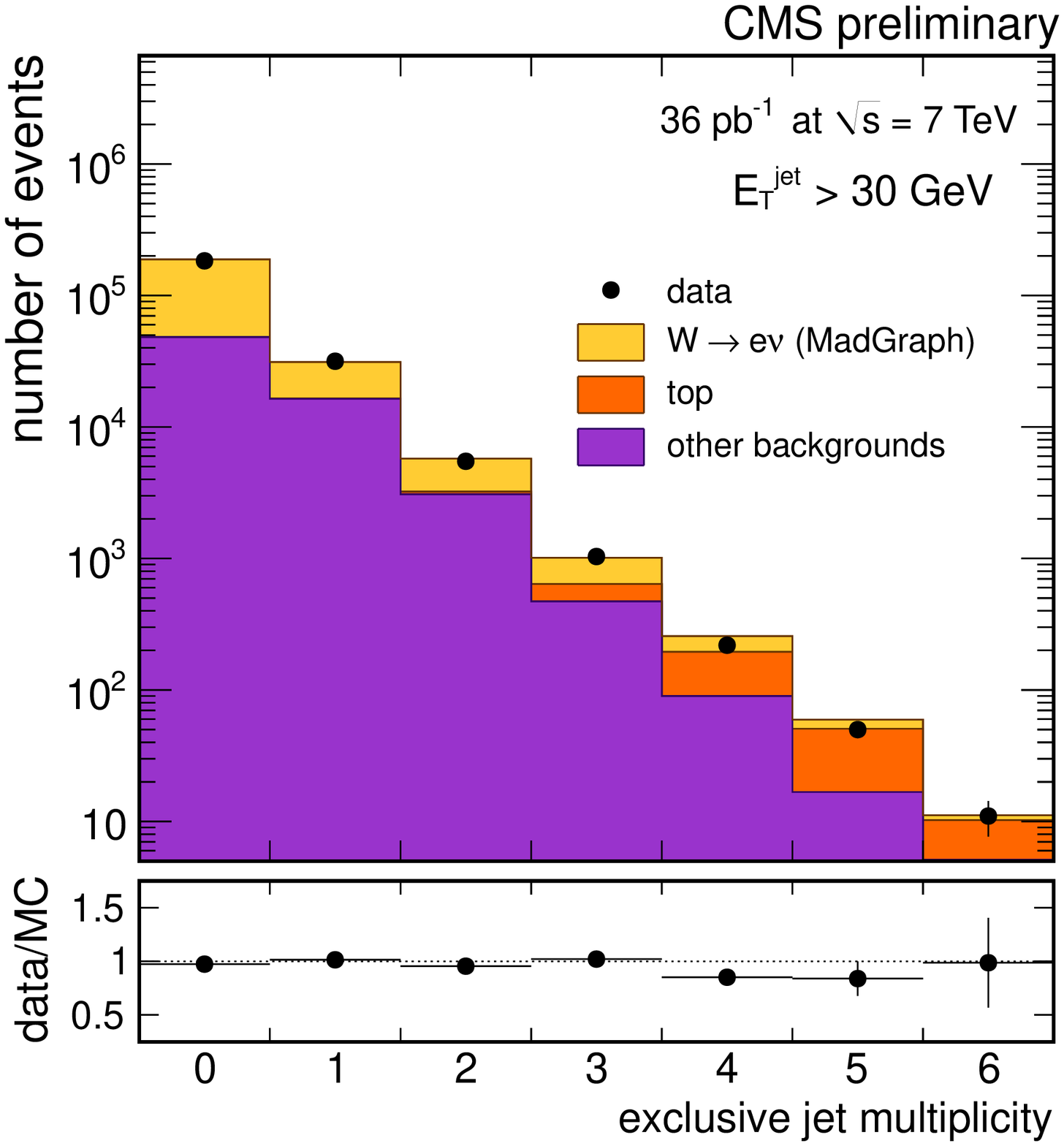}
\caption{The ATLAS uncorrected inclusive jet multiplicity distribution is shown on the left, for $W\rightarrow \mu \nu$ events. The CMS uncorrected inclusive jet multiplicity distribution is shown on the right, for for $W\rightarrow \mu e$ events. \label{fig:W_jets1} }
\end{center}
\end{figure}

The jet multiplicity distributions for $W$ production from CMS~\cite{cms_wjets}~\cite{atlas_wjets} (electron channel) and ATLAS (muon channel) are shown in Fig.~\ref{fig:W_jets1}. For high jet multiplicities, there can be significant backgrounds, especially from $t\bar{t}$ production, and special care has to be taken to minimize these backgrounds. 

The ATLAS jet transverse momenta distributions for W + jets are shown below in Fig.~\ref{fig:W_jets2}, compared to predictions from both fixed order NLO QCD (MCFM~\cite{MCFM} and Blackhat+Sherpa~\cite{blackhat}), as well as LO matrix element plus parton shower predictions.  Good agreement is observed with some indication of the theory being too large at the highest $p_T$ values. 

\begin{figure}
\begin{center}
\includegraphics[width=0.35\textwidth]{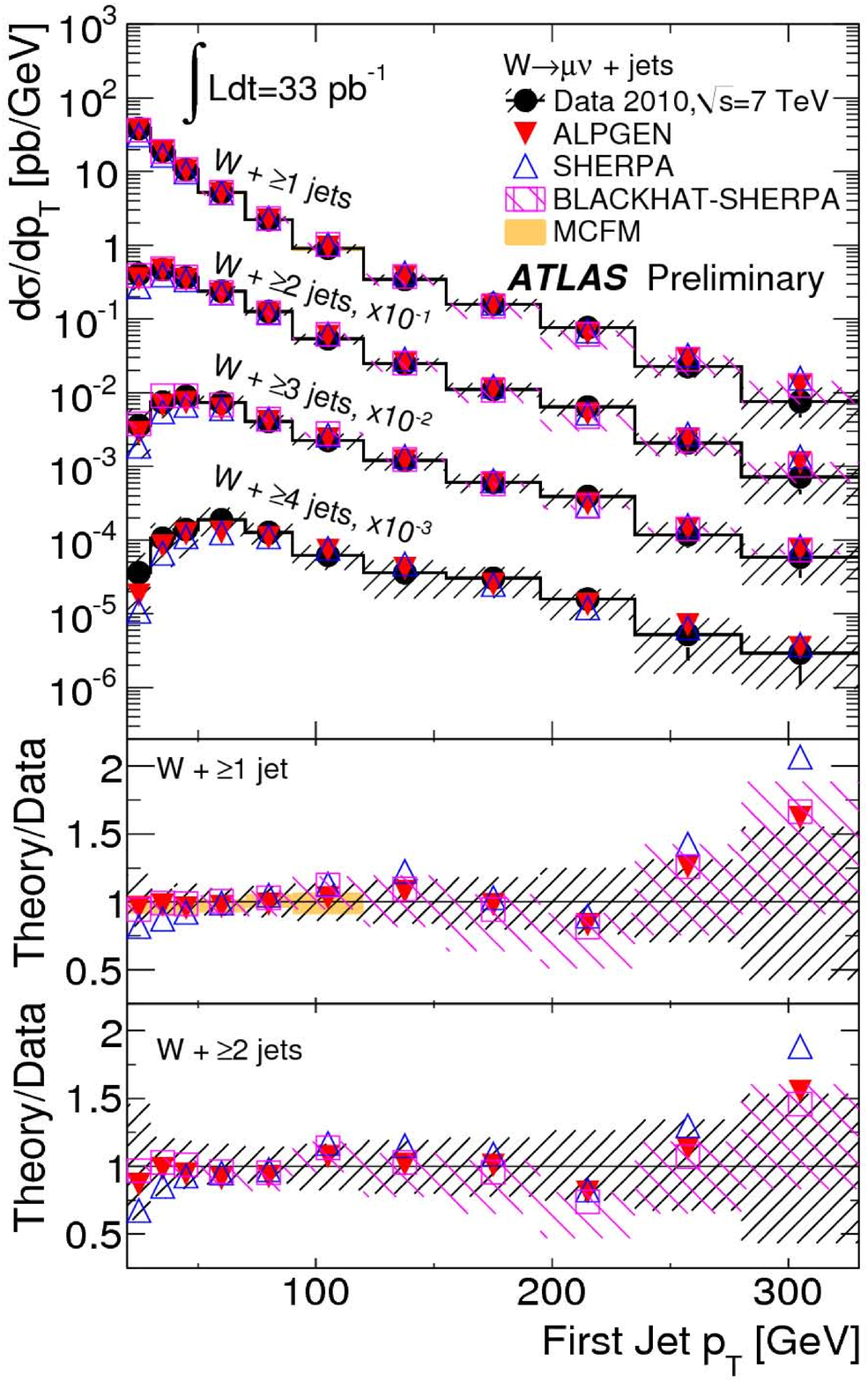}
\includegraphics[width=0.50\textwidth]{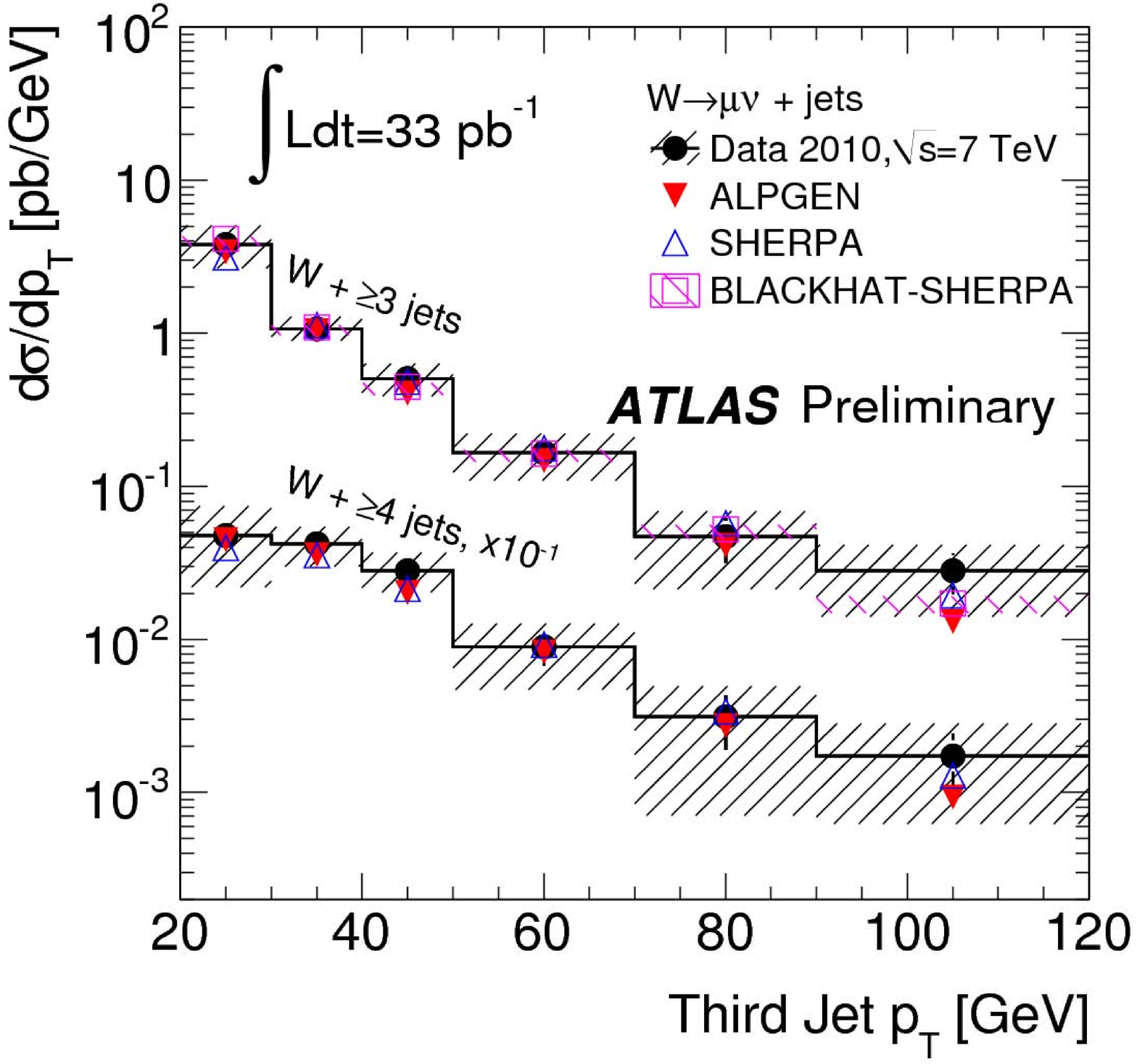}
\caption{The ATLAS  $W$ + jets cross sections, for the $W\rightarrow \mu \nu$ channel, compared to several theoretical predictions.\label{fig:W_jets2} }
\end{center}
\end{figure}

\subsection{Editorial Comment}

\begin{figure}
\begin{center}
\includegraphics[width=0.58\textwidth]{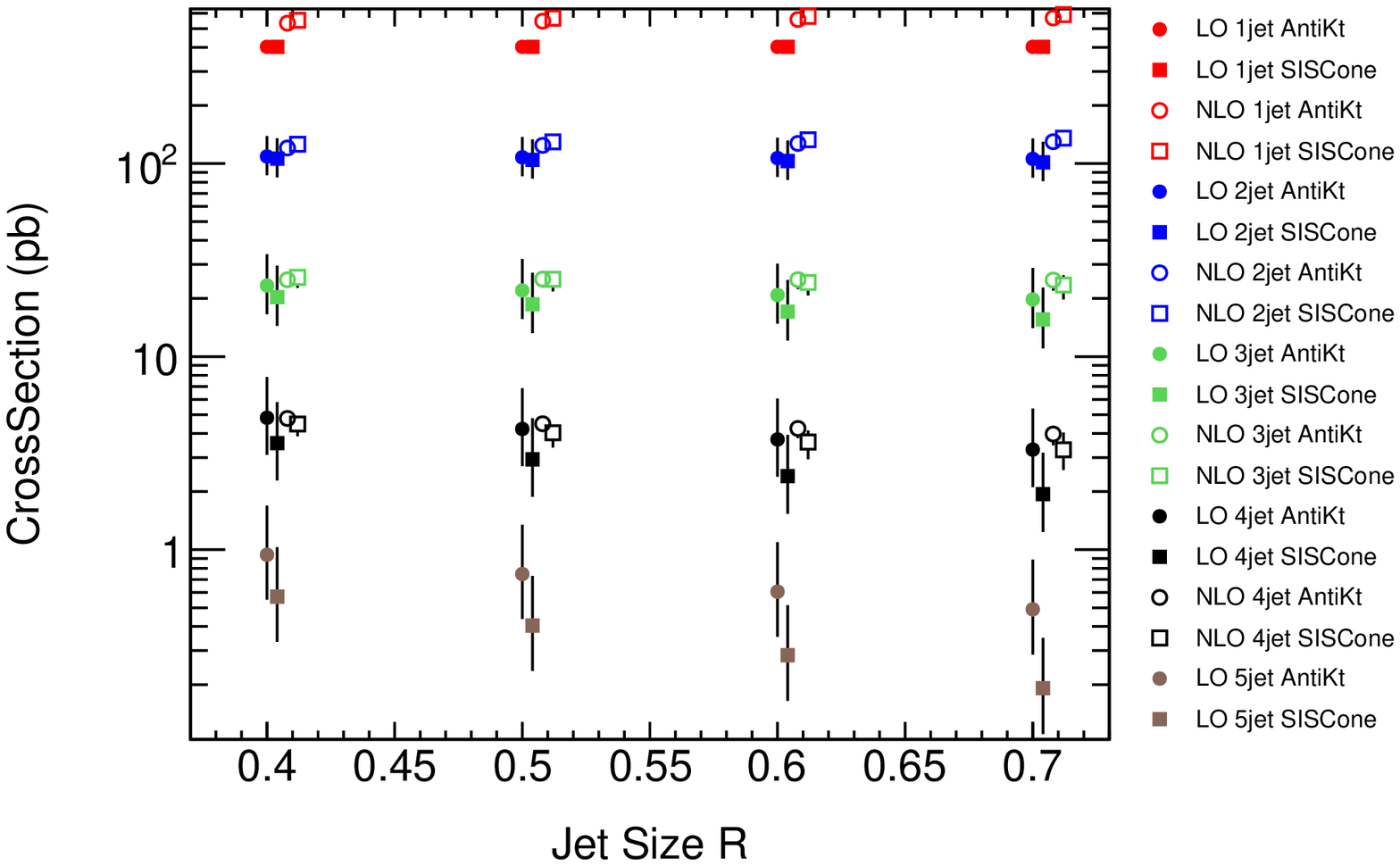}
\includegraphics[width=0.40\textwidth]{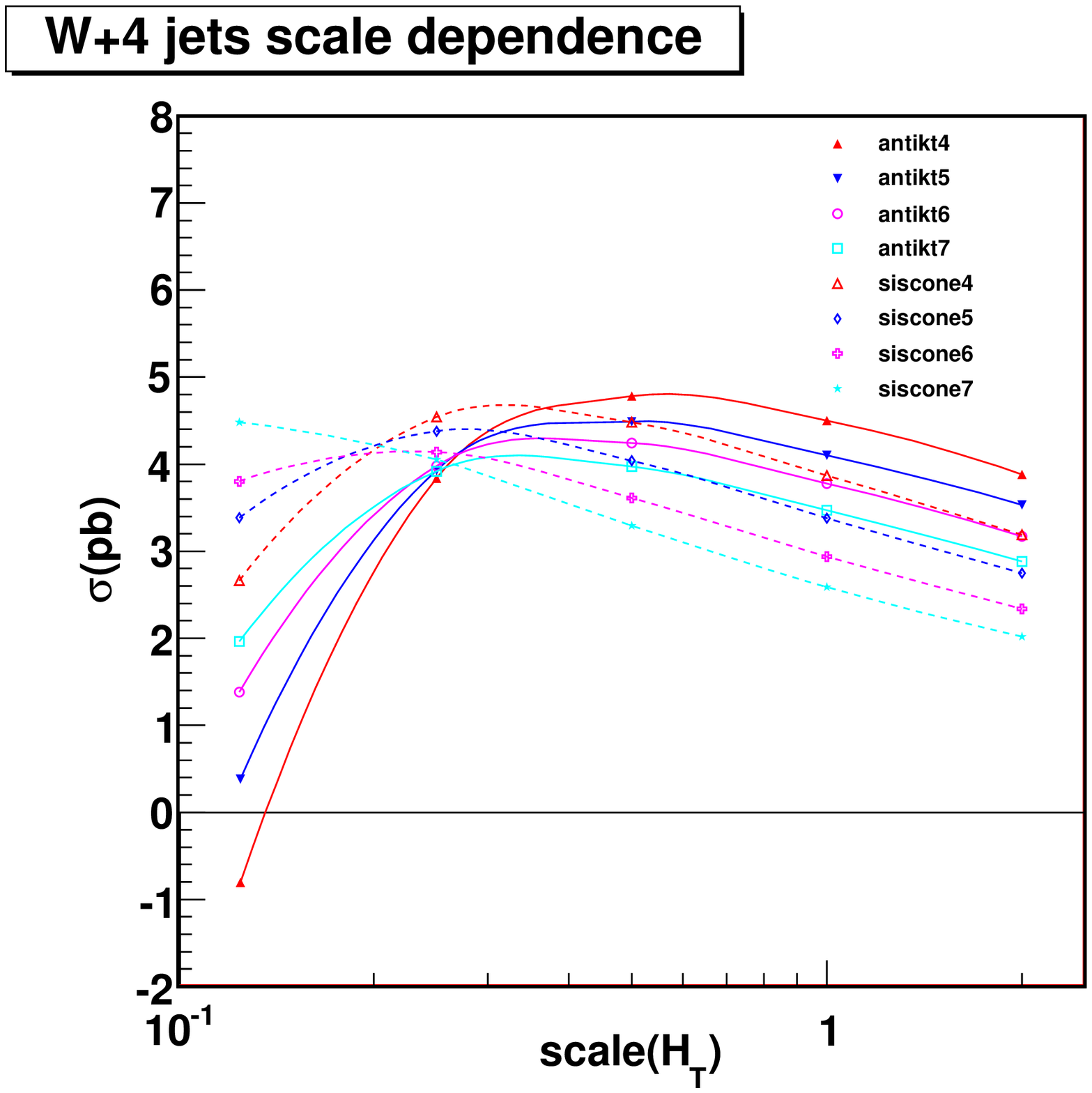}
\caption{On the left is shown the cross section times branching ratio predictions for $W$ plus 1 through 5 jets at LO and NLO, as a function of jet size, using the antikT and SISCone jet algorithms. On the right are shown the predictions for the scale dependence for $W$ + 4 jet production for 4 different jet sizes and two different jet algorithms. \label{fig:W_jets3} }
\end{center}
\end{figure}

The  ATLAS predictions shown above were the first use of Blackhat+Sherpa ROOT ntuples by experimenters. For theoretical predictions to be truly useful, they must be accessible for experimenters to allow for complete comparisons to be made. In most cases, this can be done by the use of public programs. The most optimal is the inclusion of the relevant matrix elements in a NLO Monte Carlo framework, such as MC@NLO~\cite{MC@NLO} or Powheg~\cite{Powheg}. The steps towards automatic inclusion of NLO matrix elements (aMC@NLO~\cite{aMCatNLO}) are most welcome. Some calculations, however, may be too complex (See for example some of the calculations in the Les Houches NLO wishlist~\cite{leshouches}.) to be run by non-authors, and in this case the availability of ROOT ntuples may be the best solution. Within the context of ROOT ntuple analysis, it is relatively easy to calculate the PDF/$\alpha_s(m_Z)$/scale/jet uncertainties for any observable. As an example, see Fig.~\ref{fig:W_jets3}(left), where the W + n jet cross sections (n=1-5) and their uncertainties have been calculated at NLO and LO (LO only for n=5) as a function of jet size for two different jet algorithms (anti-kT and SISCone). On the right of the figure are shown the scale dependences (with the scale varied from $0.125*H_T$ to $2.0*H_T$) for the cross sections for W + 4 jets, calculated as a function of jet size and jet algorithm. Note that the scale values for which the cross sections peak, and the shapes of the scale uncertainty curves, depend on both the jet algorithm and the jet size.  Both of these figures were produced with the Blackhat+Sherpa ntuples~\cite{blackhat}). 

\section{Inclusive photon and diphoton production}

\begin{figure}
\begin{center}
\includegraphics[width=0.40\textwidth]{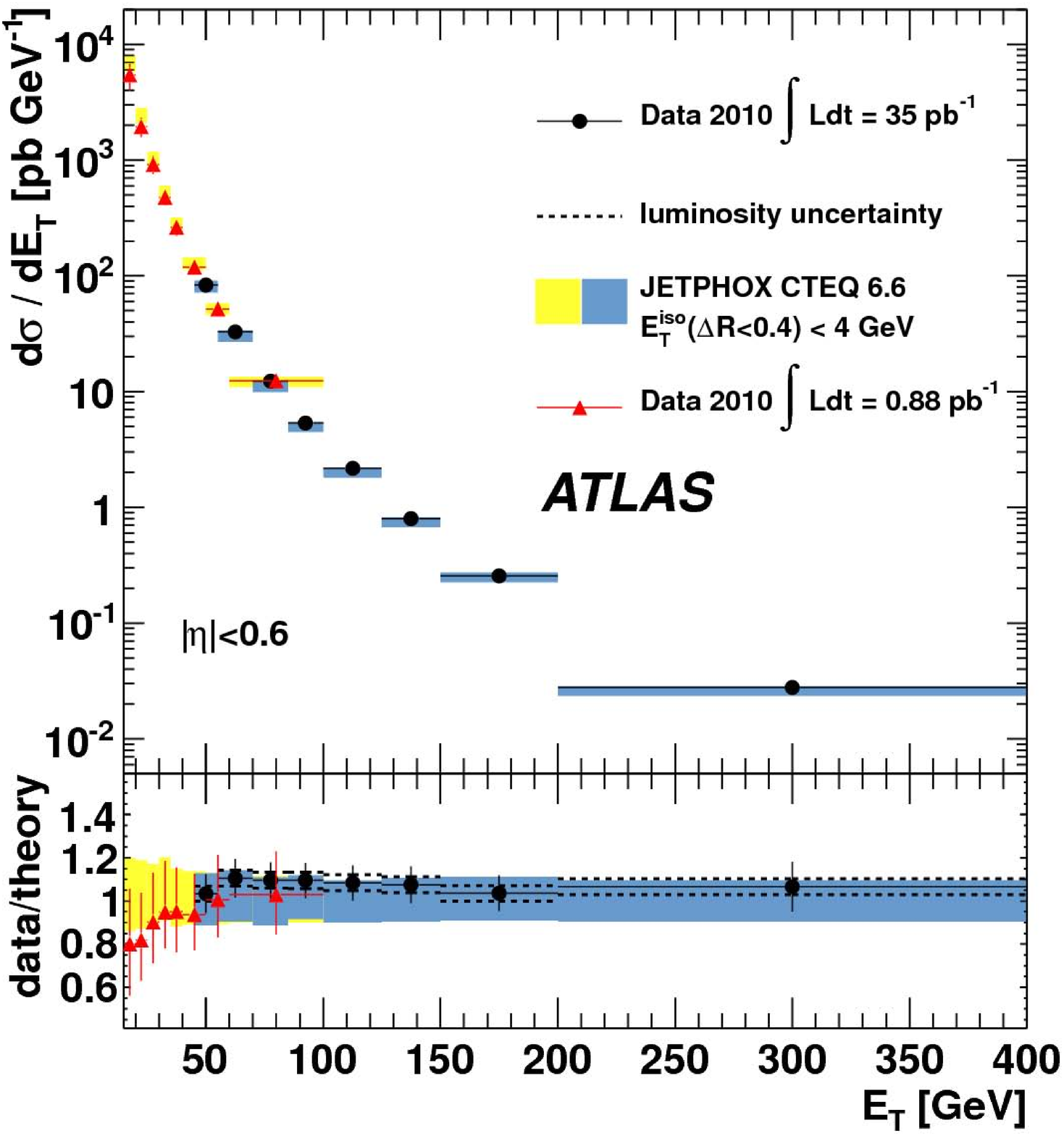}
\includegraphics[width=0.40\textwidth]{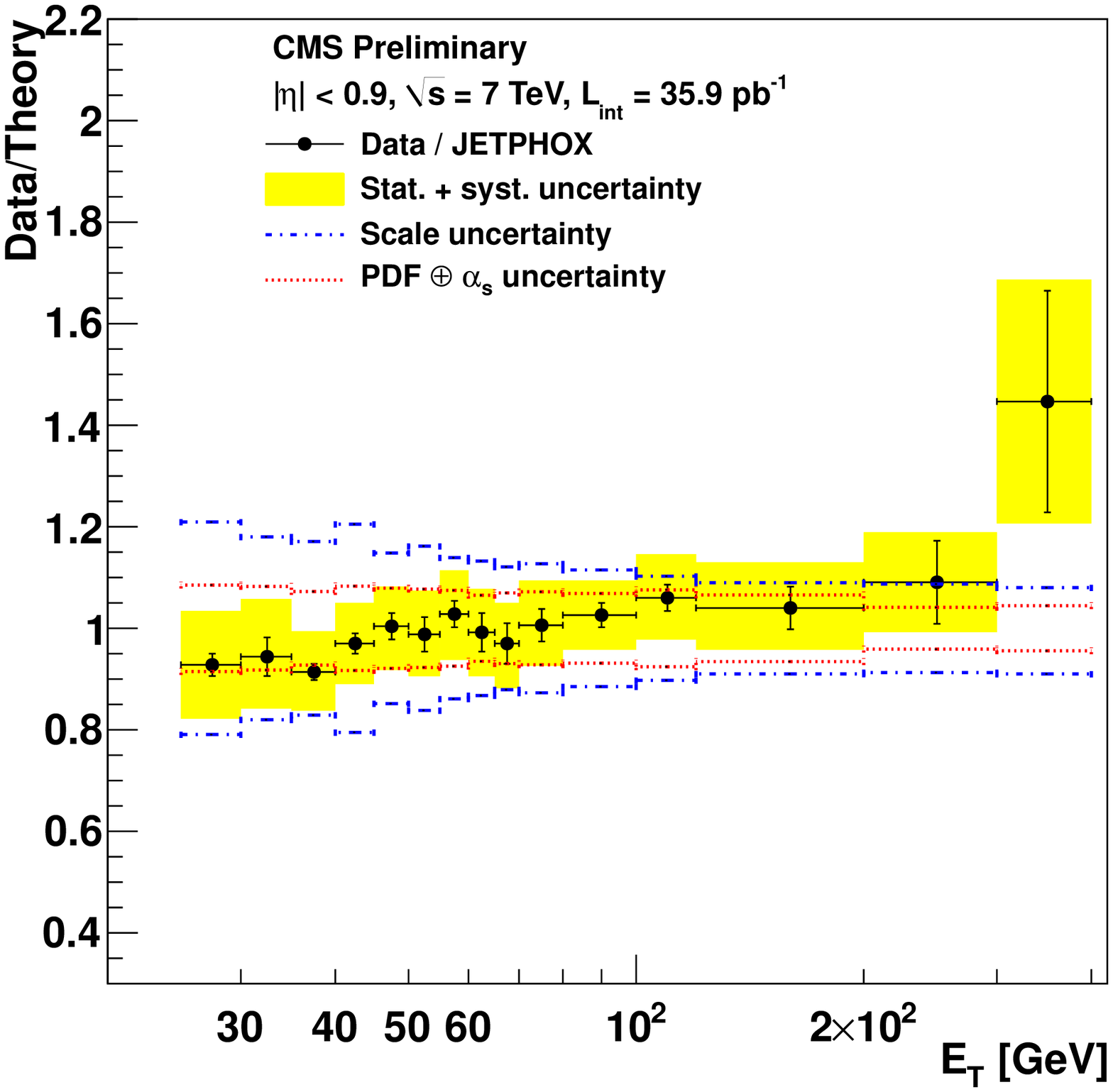}
\caption{(left) The inclusive photon cross section for ATLAS for the central rapidity region compared to the predictions from JetPhox~\cite{JetPhox} using CTEQ6.6 PDFs. (right) A linear comparison of the CMS inclusive photon cross section (for the central rapidity region) to predictions from JetPhox, using the PDF4LHC prescription for the central value and PDF uncertainty.  \label{fig:photon_ATLAS_CMS} }
\end{center}
\end{figure}

ATLAS and CMS have measured inclusive photon and diphoton production in 2010. Good agreement with the NLO predictions are observed for inclusive photon production over a wide kinematic range, with some evidence for negative deviations from theory at low $E_T$ (see Fig.~\ref{fig:photon_ATLAS_CMS}~\cite{atlas_phot,cms_phot,cms_phot2}. This is in contrast to the positive deviations observed at the Tevatron in this kinematic region. Fragmentation processes are more important at the LHC than at the Tevatron, and itÕs possible that may still be unresolved issues regarding fragmentation and isolation in theory versus experiment~\cite{leshouches}.

\begin{figure}
\begin{center}
\includegraphics[width=0.35\textwidth]{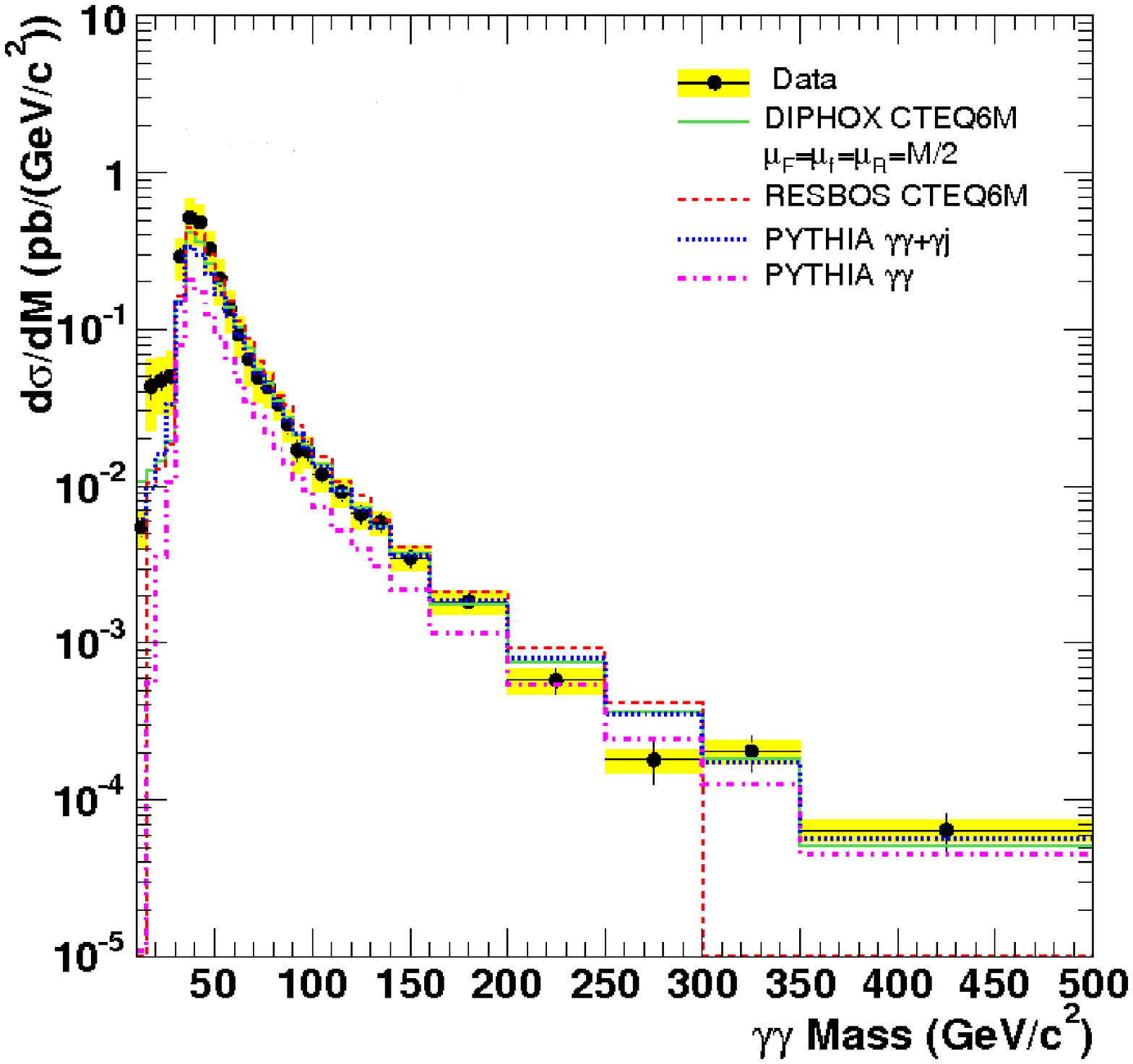}
\includegraphics[width=0.35\textwidth]{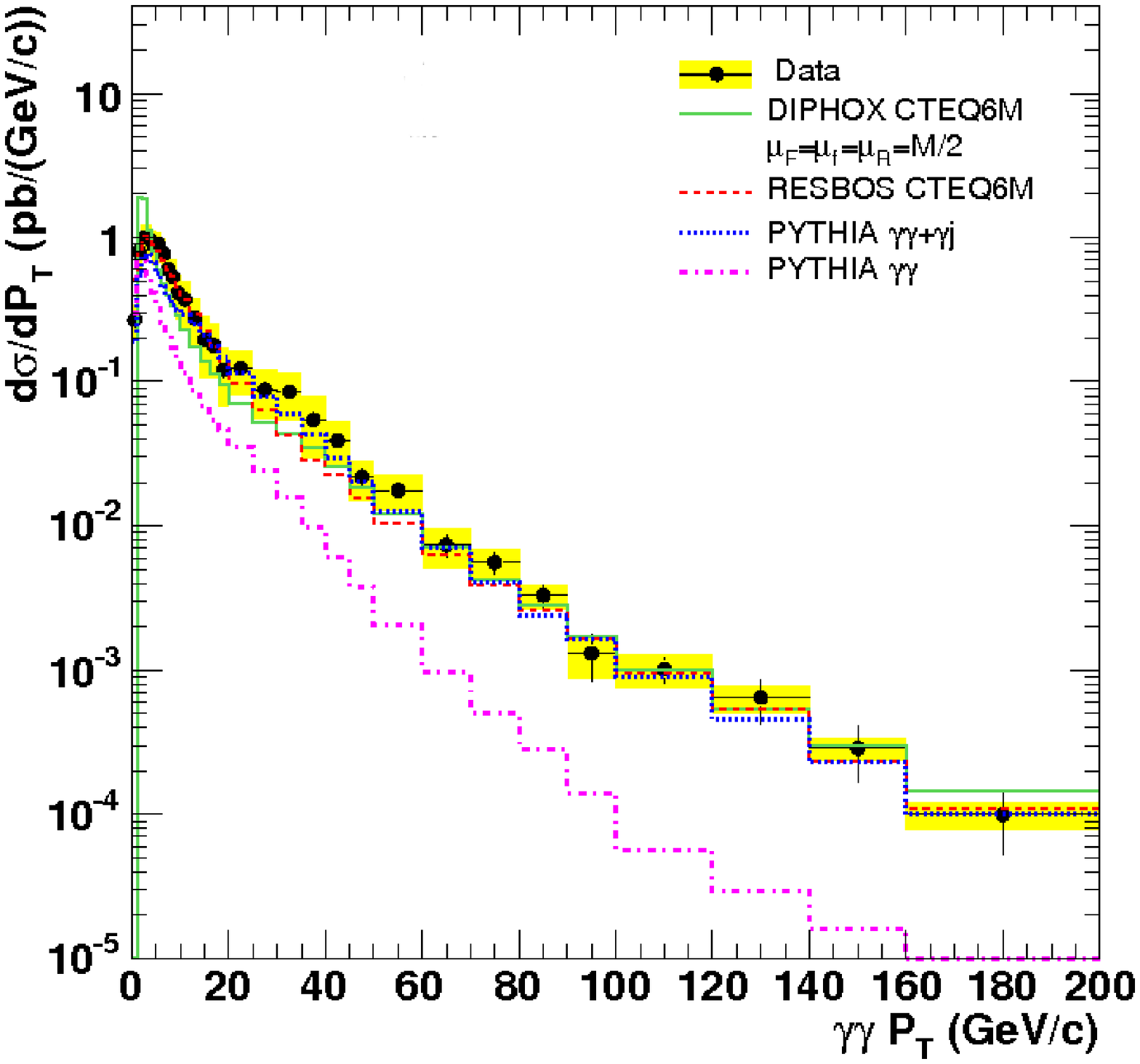}
\caption{A comparison of the CDF diphoton mass (left) and diphoton $p_T$ (right) cross sections to predictions from DiPhox, ResBos and Pythia. A cut of 17 GeV has been made for the transverse energy of the lead photon and a cut of 15 GeV has been made for the second photon.  \label{fig:cdf_diphot} }
\end{center}
\end{figure}

There have been several recent measurements of diphoton production at the Tevatron~\cite{cdfdiphot1,cdfdiphot2,d0diphot}. The CDF result is shown in Fig.~\ref{fig:cdf_diphot}, where comparisons are made to the diphoton mass, $p_T$ and $\Delta\Phi$ distributions. The data indicate  the need for soft gluon resummation at low to moderate diphoton $p_T$ and the presence of large fragmentation contributions (above those of the predictions) at low diphoton mass/small $\Delta\phi$. No prediction by itself is able to describe all kinematic regions. Somewhat surprisingly, Pythia, including fragmentation contributions of diphoton production, is able to provide reasonable agreement. 

\begin{figure}
\begin{center}
\includegraphics[width=0.35\textwidth]{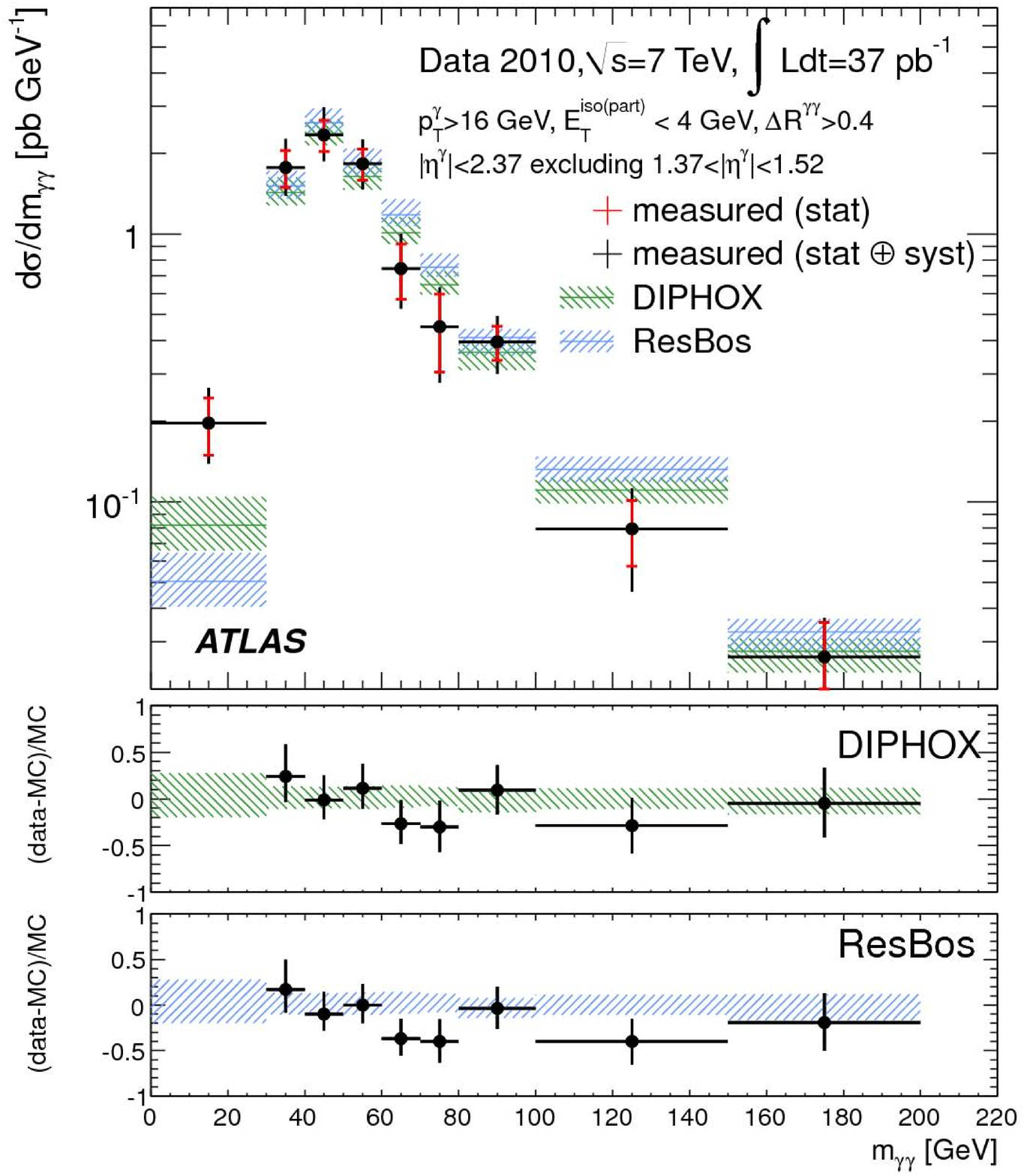}
\includegraphics[width=0.35\textwidth]{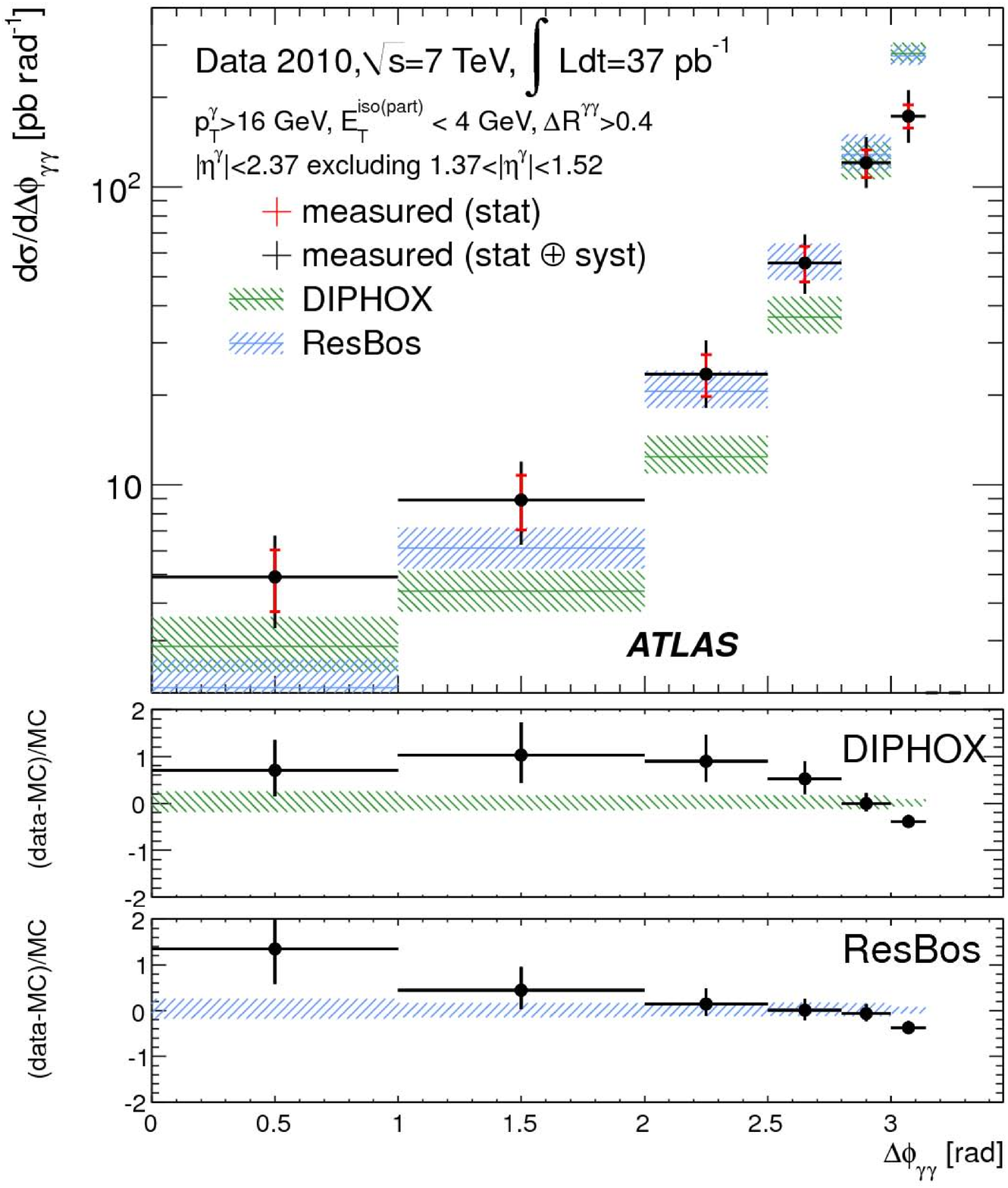}
\caption{A comparison of the ATLAS diphoton mass (left) and $\Delta \phi \phi$ cross sections to predictions from DiPhox and ResBos. \label{fig:atlas_diphot} }
\end{center}
\end{figure}

\begin{figure}
\begin{center}
\includegraphics[width=0.35\textwidth]{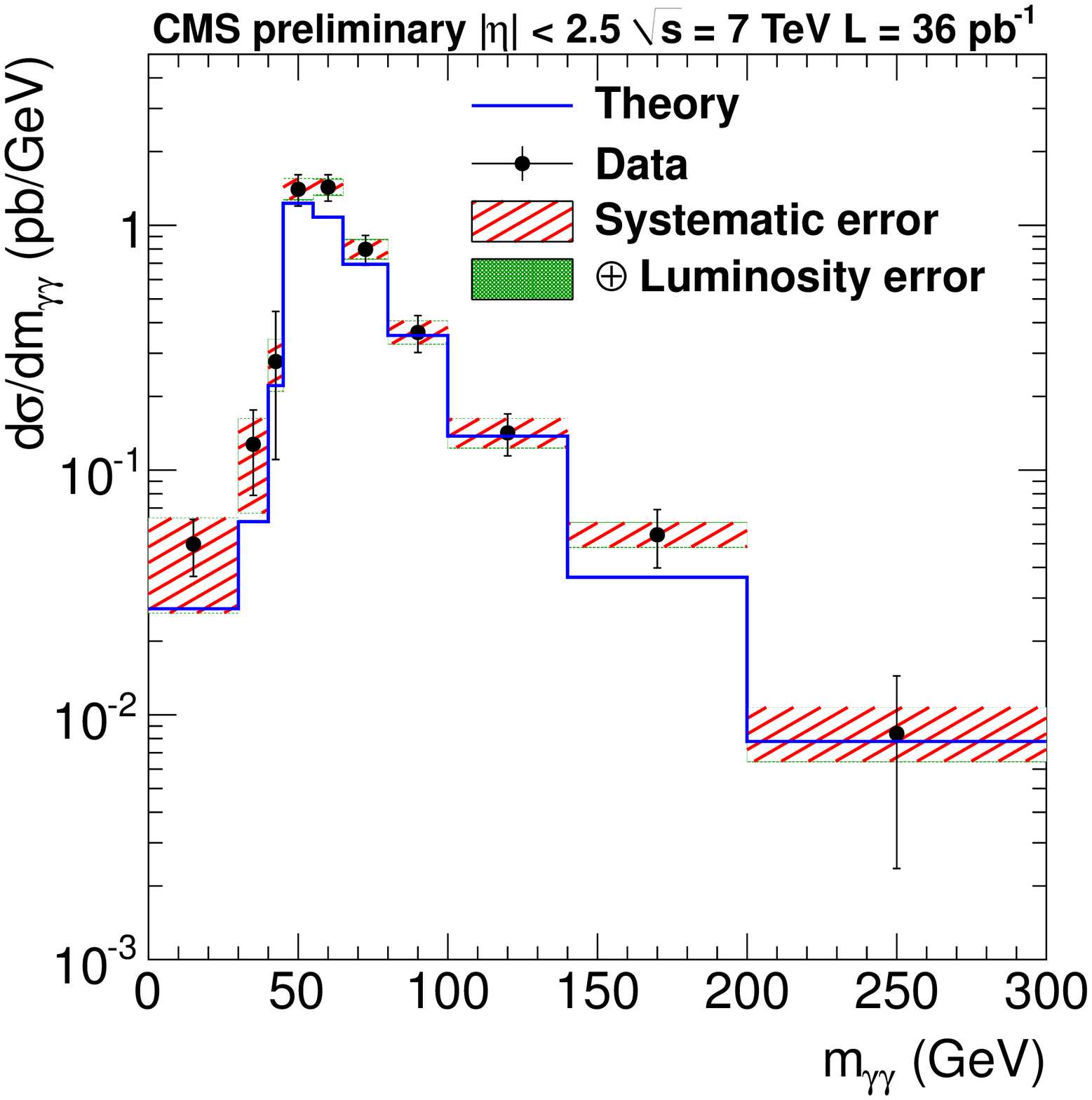}
\includegraphics[width=0.35\textwidth]{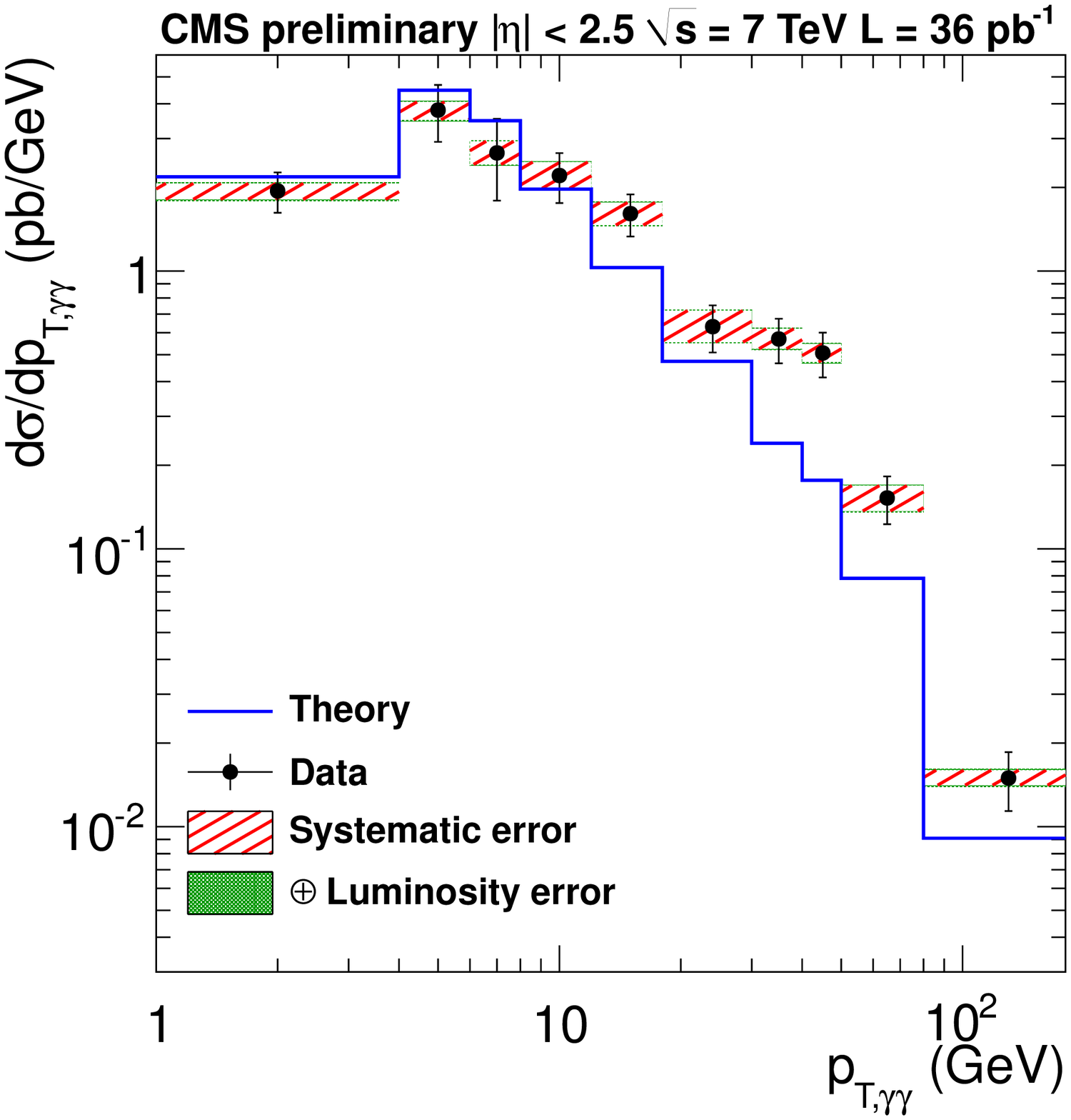}
\caption{A comparison of the CMS diphoton mass (left) and diphoton $p_T$ (right) to predictions from DiPhox. \label{fig:cms_diphot} }
\end{center}
\end{figure}

Given the importance of the diphoton decay channel for low mass Higgs production at the LHC, the measurement of inclusive diphoton production has been given a large emphasis by ATLAS and CMS. Results are shown for ATLAS for the diphoton mass and $\Delta\Phi$ distributions (Fig.~\ref{fig:atlas_diphot})~\cite{atlas_diphot}, and for CMS~\cite{cms_diphot} for the diphoton mass and  $p_T$ distributions (Fig.~\ref{fig:cms_diphot}). Similar deviations as those observed at the Tevatron are observed, indicating the need for substantial fragmentation contributions not accounted for in the existing perturbative predictions. Luckily, though, these fragmentation contributions are large primarily at low diphoton mass, and thus do not seriously impact the Higgs search region~\cite{pavel}. 

\section{Summary}

The LHC data continues to pour in, allowing for detailed comparisons to, and understanding of, perturbative QCD at the energy frontier. The data is in broad good agreement with the perturbative predictions, but there are enough open questions to keep life interesting, especially towards the kinematic edges. Recent years have seen the development of a broad array of powerful theoretical tools for analysis of and comparison to the LHC data. Added to the flexibility that exists in the experimental analysis strategies at the LHC, we have the capability of understanding the QCD environment at the LHC in far greater detail than possible at the Tevatron and at HERA.  We need to make full use of the capabilities of our detectors/analysis strategies, and of the available theory, by making use of multiple jet algorithms/sizes. All in all, this should be an interesting upcoming decade. 

\begin{acknowledgments}
I would like to thank the organizers of the conference for the invitation and for the stimulating venue. I would also like to thank Zvi Bern, Lance Dixon, Daniel Maitre, Brian Martin, Matthew Mondragon, Albert de Roeck and Nikos Varelas for useful conversations. 
\end{acknowledgments}

\bigskip 

\end{document}